\documentclass[nohyperref]{article}

\usepackage{microtype}
\usepackage{graphicx}
\usepackage{subfigure}
\usepackage{booktabs} % for professional tables
\usepackage{bm}
\usepackage{url}
\usepackage{colortbl}
\usepackage[utf8]{inputenc} % allow utf-8 input
\usepackage[T1]{fontenc}    % use 8-bit T1 fonts
\usepackage{hyperref}       % hyperlinks
\usepackage{url}            % simple URL typesetting
\usepackage{amsfonts}       % blackboard math symbols
\usepackage{nicefrac}       % compact symbols for 1/2, etc.
\usepackage{xcolor}         % colors
\usepackage{mathrsfs}
\usepackage{algorithm}
\usepackage{algorithmic}
\usepackage{tabularx}
\usepackage{multirow}
\usepackage[section]{placeins}
\usepackage{color}

\usepackage{hyperref}

\usepackage[accepted]{icml2023}

\usepackage{amsmath}
\usepackage{amssymb}
\usepackage{mathtools}
\usepackage{amsthm}

\usepackage[capitalize,noabbrev]{cleveref}

\theoremstyle{plain}

\theoremstyle{definition}

\theoremstyle{remark}

\usepackage[textsize=tiny]{todonotes}

\icmltitlerunning{Reprogramming Pretrained Language Models for Antibody Sequence Infilling}

\begin{document}

\twocolumn[
\icmltitle{Reprogramming Pretrained Language Models for Antibody Sequence Infilling}

\icmlsetsymbol{equal}{*}

\begin{icmlauthorlist}
\icmlauthor{Igor Melnyk}{ibm}
\icmlauthor{Vijil Chenthamarakshan}{ibm}
\icmlauthor{Pin-Yu Chen}{ibm}
\icmlauthor{Payel Das}{ibm}
\icmlauthor{Amit Dhurandhar}{ibm}
\icmlauthor{Inkit Padhi}{ibm}
\icmlauthor{Devleena Das}{sch}
\end{icmlauthorlist}

\icmlaffiliation{ibm}{IBM Research, Yorktown Heights, NY 10598, USA.}
\icmlaffiliation{sch}{Georgia Institute of Technology, Atlanta, GA 30332, USA. This work was done during Devleena Das's internship at IBM Research}

\icmlcorrespondingauthor{Igor Melnyk}{igor.melnyk@ibm.com}

\icmlkeywords{Machine Learning, ICML, Antibody, Infilling, CDR, Model Reprogramming}

\vskip 0.3in
]



\printAffiliationsAndNotice{}  % leave blank if no need to mention equal contribution

\begin{abstract}
Antibodies comprise the most versatile class of binding molecules, with numerous applications in biomedicine. Computational design of antibodies involves generating novel and diverse sequences, while maintaining structural consistency. Unique to antibodies, designing the complementarity-determining region (CDR), which determines the antigen binding affinity and specificity, creates its own unique challenges. 
Recent deep learning models have shown impressive results, however the limited number of known antibody sequence/structure pairs frequently leads to degraded performance, particularly lacking diversity in the generated sequences. In our work we address this challenge by leveraging Model Reprogramming (MR), which repurposes pretrained models on a source language to adapt to the tasks that are in a different language and have scarce data -- where it may be difficult to train a high-performing model from scratch or effectively fine-tune an existing pre-trained model on the specific task. Specifically, we introduce ReprogBert in which a pretrained English language model is repurposed for protein sequence infilling -- thus considers cross-language adaptation using less data. Results on  antibody design benchmarks show that our model on low-resourced antibody sequence dataset provides highly diverse CDR sequences, up to more than a two-fold increase of diversity over the baselines, without losing  structural integrity and naturalness. The generated sequences also demonstrate  enhanced antigen binding specificity and virus neutralization ability. Code is available at \url{https://github.com/IBM/ReprogBERT}
\end{abstract}

\section{Introduction}
Antibodies have emerged as essential therapeutic agents in the treatment of cancer and various other autoimmune, infectious and metabolic diseases.  Since 1985, approximately 100 monoclonal antibodies (mAbs) have been designated as drugs by FDA~\citep{jin2022emerging}. Compared to small molecule drugs, the advantage of using antibody proteins as therapeutics is their high specificity resulting in less adverse effects. A key challenge in antibody design is tailoring their binding specificity, which is  mainly influenced by the complementarity determining region (CDR). CDR plays a crucial role in antigen recognition and binding processes. It is composed of six hypervariable loops, three formed by each of heavy (H) and light (L) chains. Together, the CDRs shape  the antigen binding site of the antibody. 

Five of the six loops usually adopt well-characterized canonical conformations. In contrast, the CDR-H3 loop shows substantial variability in sequence and structure, and hence cannot be described by a canonical structure model. When compared to other protein loop structures, the CDR-H3 stands out with its significantly higher structural diversity.

\begin{figure}[!ht]
\centering
\includegraphics[width=0.8\linewidth]{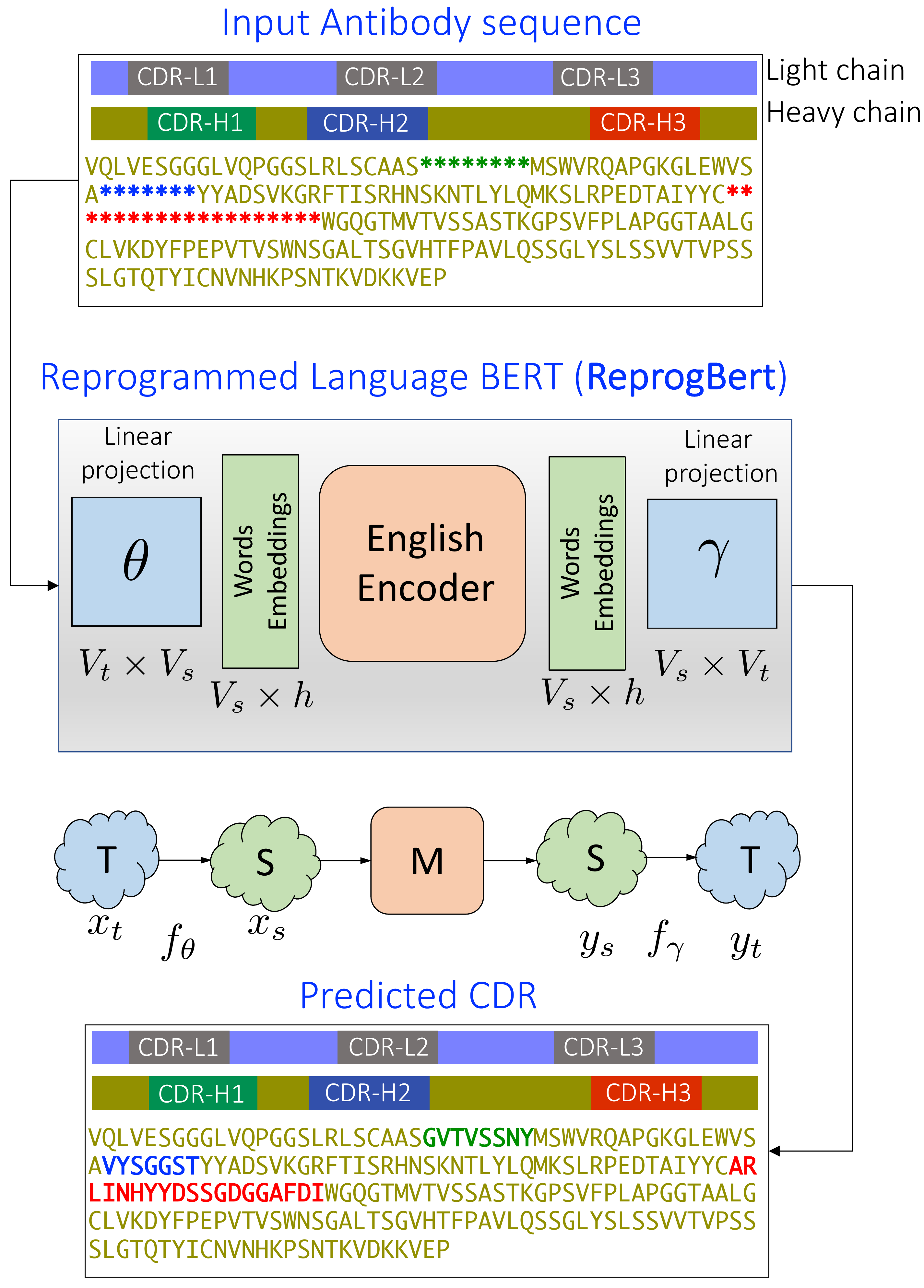}
\caption{Overview of the proposed Protein Sequence Infilling using Model Reprogramming. Given a heavy chain of an antibody, the goal is to design three Complementarity-Determining Regions (CDR-H1, CDR-H2, CDR-H3), shown in green, blue and red colors, using information from the rest of the protein. The infilling problem is formulated similar to the masked-language modeling task, where the missing amino acids are marked with a token $\mathtt{\langle MASK \rangle}$ and the model generates tokens to infill them. We emphasize that \emph{our system is a sequence-only method}, and while the structure information might be available, our method does not rely on it in the generation process. It makes the model computationally efficient while still achieving high sequence  recovery and diversity rates as compared to the current baselines. Reprogrammed language BERT model (\emph{ReprogBert}) is our proposed infilling model, where the English language BERT remains unchanged and frozen (source domain), and we introduce additional amino acid embeddings (target domain) together with the linear matrices ($\theta \in \mathbb{R}^{|V_t| \times |V_s|}$ and $\gamma \in \mathbb{R}^{|V_s| \times |V_t|}$) to project from one domain to another. During CDR infilling training, only the projection matrices and protein embeddings are fine-tuned, the language model remains unmodified. The lower diagram shows the schematic view of the reprogramming: $f_{\theta}: x_{t} \rightarrow x_{s}$ is  transforming input protein sequence (target domain (T)) into input word sequence (source domain (S)) and $g_{\gamma}: y_{s} \rightarrow y_{t}$ reverses the mapping. Thus, for a masked protein sequence $x_t$ we get predicted CDR-infilled antibody $y_t = f_\gamma(M(f_\theta(x_t)))$, where $M$ is the pretrained language model.}
\label{fig:system_overview}
\end{figure}

There is a high demand and need for efficient in-silico methods for  designing CDRs with improved specificity and other desired properties, to reduce the cost and time associate with wet lab production and testing of antibody candidates.  Generative machine learning has emerged as an attractive and viable path for this purpose. For example, for a more general task of protein design, creating new protein sequences that fold to a desired 3D structure and/or exhibit a specific function, many deep generative models
have been adapted and expanded~\citep{ingraham2019generative, cao2021fold2seq, karimi2020novo,syrlybaeva2022deep, lee2022proteinsgm, anand2022protein}. However, compared to other protein design challenges, CDR design~\citep{eguchiigantibody,shin2021proteinantibody,adolf2018rosettaantibodydesign, fu2022antibody,  kong2022conditional, Luo2022},  especially CDR-H3 design, comes with additional complexities, such as out-of-distribution generation to accommodate functional novelty. Additionally, in antibody design, sequence similarity may not reflect binding behavior. For example, in HER2 binding antibodies,  two very similar sequences (Levenshtein distance < 2) had opposing binding behavior~\citep{mason2021optimization}.  Furthermore, it is often desirable to explore new antigen binding modes, when designing antibodies for a target of interest.   Such  out-of-distribution sample generation remains challenging, particularly in a template-constrained generation scenario.

Most of the prior works compromise on the  sequence and structural diversity in generated CDRs for  high amino acid recovery and low root mean square deviation (RMSD) from ground truth structure. Moreover, the sequence-based models typically involve LLM training from scratch on NGS repertoire~\citep{olsen2022observed}, or GNN  training on a small sample of antibody sequence-structure pairs \citep{jin2021iterative}. The GNN-based models also come with a cost associated with inference, e.g., iterative design of nodes and edges in a graph via autoregressive decoding. 

\begin{table}[t]
\centering
\resizebox{0.4\textwidth}{!}{%
\begin{tabular}{@{}cccc@{}}
\toprule
 &
  Training &
  \begin{tabular}[c]{@{}c@{}}Updated\\ Parameters\end{tabular} &
  \begin{tabular}[c]{@{}c@{}}Results \\ Show\end{tabular} \\ \midrule
ProtBert &
  \begin{tabular}[c]{@{}c@{}}Task-adaptive\\ finetuning\end{tabular} &
  All &
  \begin{tabular}[c]{@{}c@{}}Accurate Recovery\\ Low Diversity\end{tabular} \\ \midrule
EnglishBert &
  \begin{tabular}[c]{@{}c@{}}Domain-adaptive\\ finetuning\end{tabular} &
  All &
  \begin{tabular}[c]{@{}c@{}}Accurate Recovery\\ Low Diversity\end{tabular} \\ \midrule
ReprogBert &
  \begin{tabular}[c]{@{}c@{}}Language-adaptive\\ reprogramming\end{tabular} &
  {$\pmb{\theta}$, $\pmb{\gamma}$ \bf{only}} &
  \begin{tabular}[c]{@{}c@{}}Accurate Recovery\\ \bf{High Diversity}\end{tabular} \\ \bottomrule
\end{tabular}%
}
\caption{Comparison of our proposed methods. ReprogBert stands out as an efficient cross-language approach generating accurate and diverse protein sequences. }
\label{tab:comp}
\end{table}

To address these challenges, we propose an alternative sequence-only framework (see Fig.~\ref{fig:system_overview} for an overview), that is reprogramming existing out-of-domain English language BERT model \citep{devlin2018bert} toward the protein infilling task, given the rest of the sequence as a template. We term this model \emph{ReprogBert}. Additionally, for our sequence-based infilling task we also consider in-domain specialized protein model \emph{ProtBert} \citep{Elnaggar2020} as well as the English language BERT (\emph{EnglishBert}), whose out-of-domain language token embeddings are replaced with in-domain amino acid embeddings (see Table~\ref{tab:comp} and Fig.~\ref{fig:baselines_overview} for details). We compare all proposed infilling methods with physics-based and graph-based generative models on a list of tasks ranging from template-constrained CDR design to CDR sequences with predicted SARS-COV-2 neutralization ability. We show that while ReprogBert enjoys comparable high structural consistency, and lower sequence perplexity, when matched against the baselines, it shows high amino acid CDR recovery while providing additional benefit regarding generating highly diverse CDR sequences. These results suggest the potential of ReprogBert toward on-demand generation of the out-of-distribution sequences in the learning from limited data scenario. The other proposed baseline systems, EnglishBert and ProtBert, achieve high CDR sequence recovery rates with consistent structural integrity, although having a modest sequence diversity performance.

In summary, in this work we: \textbf{(i)} propose ReprogBert, a system for protein sequence infilling using model reprogramming for the task of antibody CDR design,  \textbf{(ii)} show promising performance results as compared to many baselines (including our own proposed ProtBert and EnglishBert baseline infilling methods) and over multiple benchmarks, where our ReprogBert model upholds structural integrity and sequence recovery, while achieving valuable high diversity of the generated sequences. Moreover, the generated CDR sequences frequently have the lowest perplexity, reflecting their well-formed composition and naturalness. ReprogBert further shows its promise in harder CDR design tasks, can handle multiple CDR infilling at once, and does not need structure template information,   and \textbf{(iii)} observe high data-efficiency of the reprogrammed model, having only a few training parameters, it can be efficiently trained in the data-scarce domains, such as antibody design, while still leveraging information from large out-of-domain language pretraining.

\section{Reprogramming for Protein Sequence Infilling}
\label{sec:infill}

The field of model reprogramming (MR) has focused on repurposing pretrained machine learning (ML) models for varied ML tasks in different domains. It was firstly proposed in an adversarial setting of stealthy resource alternation in \citep{elsayed2018adversarial} and later extended to cross-domain resource-efficient transfer learning \citep{chen2022model,neekhara2022cross}. MR achieves state-of-the-art performance in many tasks, especially in data-limited settings, including reprogramming general images for bio-medical measurements \citep{tsai2020transfer}, human voice for time-series \citep{yang2021voice2series}, and sentence sentiment for protein property \citep{vinod2020reprogramming}. 
While current MR techniques focus on classification tasks, in our work we seek to extend MR capabilities into generative tasks through reprogramming large pretrained language models for protein sequence infilling. To the best of our knowledge, this work is the first study for such an  endeavor. 

We also note that the term reprogramming %is used in the literature in a sense which 
is quite different from fine-tuning. Finetuning aims for task-related (e.g., finetuning a ProtBert model on antibody design task) or domain-specific (e.g., adapting an EnglishBert to the medical domain) adaptation of the pretrained model, while the goal of reprogramming is cross-language adaptation, i.e. English to protein \cite{ruder2021lmfine-tuning}.  %In fine-tuning, the modality/domain of the pretrained model (e.g, English/Language) and the fine-tuned model (also English/Language) is the same. 
Further, in finetuning, all of the parameters of the pre-trained models are updated~\cite{howard-ruder-2018-universal}, or task-specific layers are injected into the model \cite{adaptor}, while in reprogramming the pre-trained model remains frozen. %Reprogramming is also different from prompt engineering, as the frozen pre-trained model does not need any hand-engineered or searched prompt composed of discrete words in the same language to be fed \cite{brown2020language, shin-etal-2020-autoprompt}. 
Our approach shares some similarity with prompt-tuning~\cite{prompt-tuning, prefix-tuning, hambardzumyan-etal-2021-warp}, in which soft prompts composed of continuous embeddings are learned in an end-to-end manner, where the pre-trained model remains frozen. However, so far prompt-tuning is limited to task-specific adaptation of a pre-trained model where the task is in the same language, and the output is also in the source language, whereas reprogramming focuses on cross-language adaptation and outputs in the target language domain.

Given a protein sequence, we propose novel CDR loop design as a form of a template-infilling. The template is provided by the amino acid sequence of the constant region of the antibody, as those are conserved and less likely to change, while the sequences corresponding to CDR can vary and change the structure of the antigen binding interface, resulting into modified antigen affinity and specificity.  It should be mentioned though the infilling here is performed to design CDRs of antibodies, the framework can be leveraged to infill any protein sequences.

Figure \ref{fig:system_overview} presents an overview of our proposed framework, \emph{ReprogBert}, the reprogrammed language model for protein sequence infilling. Specifically, we use the pretrained English BERT model \citep{devlin2018bert} (in our experiments, it is the $\mathtt{base}$-$\mathtt{bert}$-$\mathtt{uncased}$ from HuggingFace) and reprogram it for infilling the CDR part of the antibodies.

The number of tokens in the original language task (i.e., source domain) is denoted by $V_s$ (in our experiments $|V_s| = 30522$ word tokens). The language sentence can then be represented as 
\begin{align}
y_{s} = \langle w_{1},w_{2},\ldots, w_{n} \rangle,
\end{align}
where $w_{i}$ is the word token. The number of tokens in the task of interest (i.e., target domain) is denoted by $V_t$ (in our experiments $|V_t| = 30$ protein tokens: 20 amino acid tokens and 10 auxiliary tokens). The protein sentence can then be represented as 
\begin{align}
x_{t} = \langle a_{1},a_{2},\ldots, a_{n} \rangle,
\end{align}
where $a_{i}$ is an amino acid token. We define two mappings (see bottom plot in Fig.~\ref{fig:system_overview}) 
\begin{align}
f_{\theta}: x_{t} \rightarrow x_{s},
\end{align}
transforming input protein sequence into input word sequence and 
\begin{align}
g_{\gamma}: y_{s} \rightarrow y_{t},
\end{align}
reversing the transformation by mapping output word sequence into protein one. Following the approach in \citep{elsayed2018adversarial, tsai2020transfer, vinod2020reprogramming} we constrain these mappings to be linear transformations between the source and target domains. Formally, these mappings are represented as
\begin{align}
x_{s} &= x_{t}\theta\\
y_{t} &= y_{s}\gamma,
\end{align}
where the linear projection matrices 
\begin{align}
\theta &\in \mathbb{R}^{|V_t| \times |V_s|}\\
\gamma &\in \mathbb{R}^{|V_s| \times |V_t|}
\end{align}
are the parameters of the transformations. In particular, focusing on the input projection matrix $\theta$ and treating $x_t$ as a one-hot sequence representation of the amino acids of length $N$, i.e., $x_t \in \mathbb{R}^{N\times |V_t|}$, the sequence representation in the English token domain becomes $x_s \in \mathbb{R}^{N\times |V_s|}$. This representation is then projected onto the embedding matrix of the English Bert model $E \in \mathbb{R}^{|V_s|\times d}$ ($d$ is the latent model dimension):
\begin{align}
x_s^{E} = x_s E,
\label{eq:Eproj}
\end{align}
and continue the usual processing through the transformer layers and blocks. 

During training, all model parameters are fixed and only $\theta$ and $\gamma$ are optimized. Specifically, we update $\theta$ and $\gamma$ with respect to minimizing $\mathcal{L}_{NLL}(y_{t}, y_{t}^*)$, the loss between the estimated infilled protein sequence $y_{t} = f_\gamma(M(f_\theta(x_t)))$, given the CDR-masked anitbody $x_t$ and the ground truth sequence $y_{t}^*$.

Note that from \eqref{eq:Eproj}, the amino acid embeddings $E_{aa} \in \mathbb{R}^{|V_t|\times d}$ can be defined as $E_{aa} = \theta E$.
In Section \ref{sec:apendix_proj} of appendix we visualize these embeddings and examine their clustering based on biological properties such as electrical charge, hydrophobicity, size, etc. (see Fig.~\ref{fig:clust4} in Appendix). We also note that since $x_s$ in equation \eqref{eq:Eproj} is usually a dense vector, where all rows of $E$ are mixed during the projection. 
%its projection onto embedding vector has limited interpretability in that . 
To better understand the mapping between amino acids and English tokens, in Section \ref{sec:apendix_crs_atn} of appendix we replace the projection with cross-attention mechanism and examine its attention weights (see Fig.~\ref{fig:map4} in Appendix).

\section{Experiments}

In this section we present evaluation results of our proposed methods on template constrained CDR design using Structural Antibody Database (SabDab)  \citep{gkt1043} and Rosetta Antibody Design (RabD) \citep{jin2021iterative}, and CoV-AbDab dataset \citep{Raybould2021} neutralization using the model's generated antibodies. %In what follows, we first discuss the evaluation metrics, followed by the introduction of the baseline models and the presentation of the results on three datasets.

\begin{figure}[!t]
\centering
\includegraphics[width=0.99\linewidth]{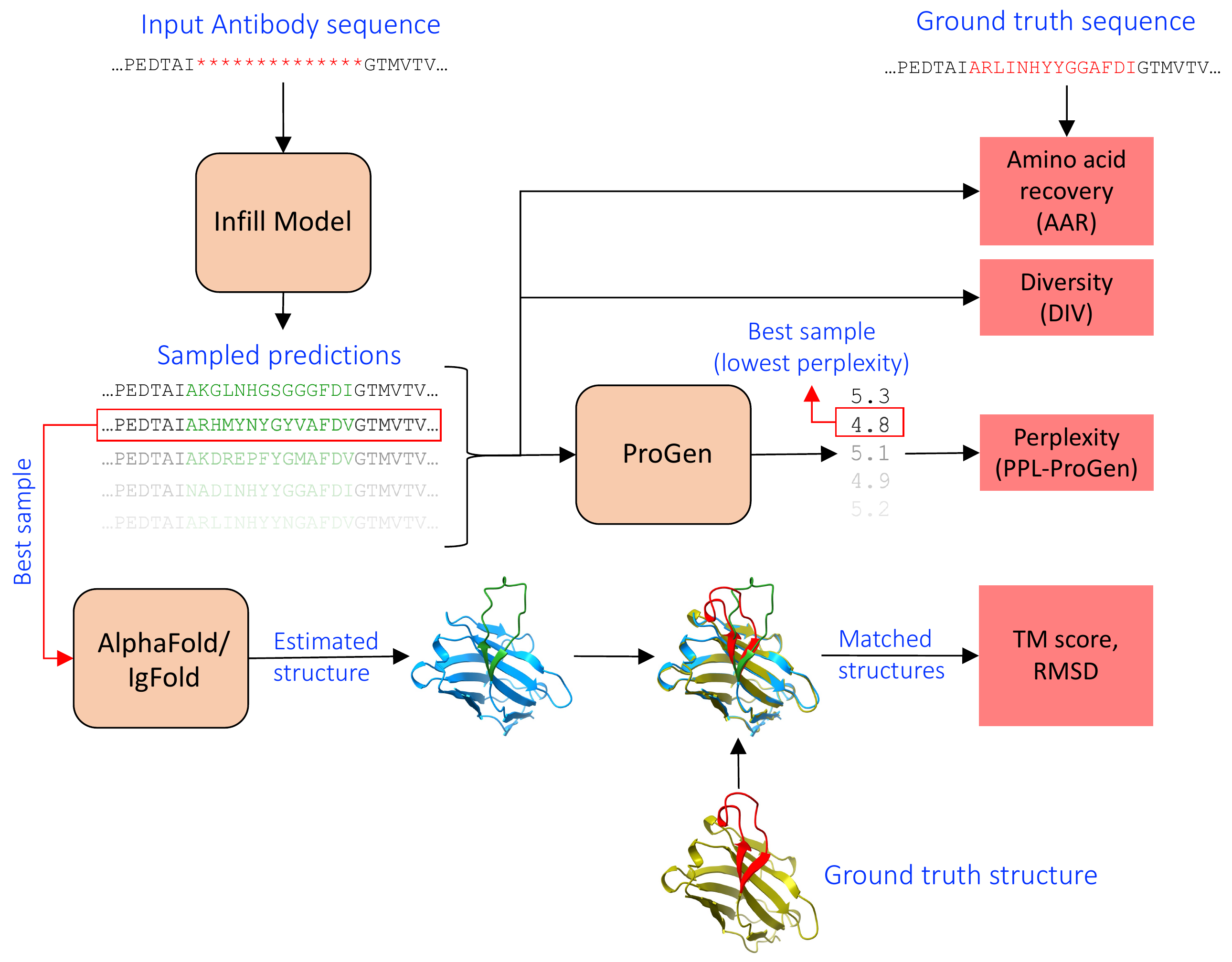}
\caption{Evaluation process and the computed metrics. For each masked  antibody input sequence we generate 100 predicted samples. Metrics are Amino acid recovery (AAR);  %is computed for the specific sequence region (e.g., CDR-H3), measuring the fraction of exact matches between ground truth and the sampled sequences. 
Diversity (DIV) --  %only the generated samples to compute the complement of the average recovery for all pairwise comparisons in the set (the higher the number, 
 dissimilarity of each sample to all the others; Perplexity (PPL-ProGen)  computed %as the average of all the sampled sequences (masking only the region of interest), 
 using off-the-shelf %autoregressive Transformer protein model 
 ProGen model \citep{nijkamp2022progen2},  reflecting ``naturalness'' of the designed sequences. The sample with the minimum perplexity (red box with an arrow) is then used for 3D structure prediction using AlphaFold~\citep{jumper2021highly} or IgFold~\citep{ruffolo2022fast} models and compared with ground truth to compute template modeling (TM) score \citep{zhang2004scoring} and the root mean squared deviation (RMSD) from the input structure.}
\label{fig:metrics}
\end{figure}

\subsection{Evaluation Metrics}
For each input protein sequence in our experiments we generated 100 samples using our infill models. To measure the quality of these samples, we then 
%follow the standard practice and 
compute the following evaluation metrics (see Fig.~\ref{fig:metrics} for an illustration). \emph{Amino acid recovery (AAR)} is computed for the specific sequence region of interest (e.g., CDR-H3), measuring the percent of the exact matches between ground truth and the sampled sequences. The range is 0-100, and the higher the AAR, the more accurate the recovery.   \emph{Diversity (DIV)}, on the other hand, uses only the sampled proteins to compute the complement of the average recovery of all pairwise comparisons in the set. Here the range is 0-100 and the higher the number, the more dissimilar are samples among themselves. While in general it holds true that the recovery and diversity are inversely correlated, i.e., higher recovery rate leads to lower diversity, and vice versa, CDR design calls for  generative models that achieve at least above 30\%  recovery \citep{weitzner2015origin}, while at the same time are able to maintain high  sequence diversity. 

\subsection{Baseline Models}

\begin{table}[!t]
\centering
\resizebox{0.45\textwidth}{!}{%
\begin{tabular}{@{}cccccc@{}}
\toprule
CDR    & Train & Validation & Test & Average CDR length & Average CDR diversity \\ \midrule
CDR-H1 & 4050  & 359        & 326 & 8.1 & 60.8\\
CDR-H2 & 3876  & 483        & 376 & 7.9 & 68.2 \\
CDR-H3 & 3896  & 403        & 437 & 14.5 & 76.9 \\ \bottomrule
\end{tabular}%
}
\caption{Statistics of the Structural Antibody Database (SabDab) for the training, validation and test splits across the three CDRs. We also show the average number of amino acids per CDR and average CDR diversity (length-normalized) across proteins. As can be seen CDR-H3 is the longest and most diverse and therefore represents the most challenging prediction task.}
\label{tab:sabdab_stats}
\end{table}

For \emph{perplexity} (the model’s predicted probabilities for every residue in a given sequence) we use off-the-shelf autoregressive Transformer protein model ProGen \citep{nijkamp2022progen2} to compute \emph{PPL-ProGen} as the average of 100 samples (masking only the region of interest). Specifically, we used ProGen2-small (151M parameters), which has been pretrained on the mixture of Uniref90 \citep{suzek2015uniref} and BFD30 \citep{steinegger2018clustering} datasets. For perplexity, the lower values mean the better performance, indicating stronger ``naturalness'' of generated CDRs. The sampled protein sequence with the minimum perplexity is then used for 3D structure prediction using protein folding model (e.g., AlphaFold \citep{jumper2021highly} or IgFold \citep{ruffolo2022fast}). The full predicted and ground truth structures are then compared to compute \emph{template modeling (TM)} score \citep{zhang2004scoring} (range 0-100, higher the better) and the \emph{root mean squared deviation (RMSD)} (lower the better), focusing only on the CDR part. The suffix AF correpsonds to AlphaFold, while IF means IgFold.

\begin{table*}[!ht]
\centering
\resizebox{0.7\textwidth}{!}{%
\begin{tabular}{@{}lccccccccc@{}}
\toprule
                  & \multicolumn{9}{c}{SabDab CDR-H1}\\
                    \cmidrule(lr){2-10} 
                  & PPL & PPL-ProGen & RMSD & RMSD-AF & RMSD-IF & TM-AF & TM-IF & AAR & DIV\\
\midrule
LSTM             & 6.79  & -- & --    & -- & -- & -- & -- & -- & -- \\
AR-GNN           & 6.47  & -- & 2.97  & -- & -- & -- & -- &  --   & --  \\
Refine-GNN       & 6.09  & \cellcolor[HTML]{EFEFEF}3.5 & 1.18  & 4.42 & 1.78 & 84.0 & 93.6 & 61.2 & \cellcolor[HTML]{C0C0C0}47.3  \\
AbLang           & --    & -- & --  & -- & -- & -- & -- & 47.7 & 42.8  \\
ProtBert         & --    & \cellcolor[HTML]{EFEFEF}3.5 & -- & 4.16 & 1.68 & 84.4 & 93.8 & \cellcolor[HTML]{C0C0C0}64.7 & 4.6 \\
EnglishBert      & --    & 3.7 & -- & 4.22 & 1.67 & 84.1 & 93.8 & \cellcolor[HTML]{EFEFEF}63.6 & 5.8 \\
ReprogBert       & --    & \cellcolor[HTML]{C0C0C0}3.3 & -- & 4.31 & 1.73 & 84.0 & 93.7 & 56.0 & \cellcolor[HTML]{EFEFEF}29.1 \\
\bottomrule
\end{tabular}%\
}
\caption{Evaluation results on the SabDab dataset for CDR-H1 in the heavy chain. Dark grey cell denote best results, while light grey are the second best. ReprogBert generates sequences with lowest perplexity, second best diversity and high AAR with structural consistency.}
\label{tab:sabdab1}
\end{table*}

\begin{table*}[!t]
\centering
\resizebox{0.7\textwidth}{!}{%
\begin{tabular}{@{}lccccccccc@{}}
\toprule
                  & \multicolumn{8}{c}{SabDab CDR-H2}\\
                    \cmidrule(lr){2-10} 
                  & PPL & PPL-ProGen & RMSD & RMSD-AF & RMSD-IF & TM-AF & TM-IF & AAR & DIV\\
\midrule
LSTM             & 7.21   & --    & --   & -- & -- & -- & -- & --  & -- \\
AR-GNN           & 6.86   & --    & 2.27 & -- & -- & -- & -- & --  & --  \\
Refine-GNN       & 6.58  & \cellcolor[HTML]{C0C0C0}3.4 & 0.87 & 3.05 & 1.40 & 85.7 & 93.9 &  48.9 & \cellcolor[HTML]{EFEFEF}38.7  \\
AbLang       & --  & -- & -- & -- & -- & -- & -- &  46.7 & \cellcolor[HTML]{C0C0C0}44.9  \\
ProtBert         & --    & \cellcolor[HTML]{EFEFEF}3.6    & --   & 3.10 & 1.32 & 85.6 & 93.9 & \cellcolor[HTML]{C0C0C0}59.5 & 5.5 \\
EnglishBert      & --    & 4.0    & --   & 3.07 & 1.32 & 85.6 & 93.9  & \cellcolor[HTML]{EFEFEF}59.1 & 7.7 \\
ReprogBert       & --    & 3.9    & --   & 3.02 & 1.40 & 85.8 & 93.8 & 53.0 & 37.9 \\
\bottomrule
\end{tabular}%
}
\caption{Evaluation results on the SabDab dataset for CDR-H2 in the heavy chain. As compared to Table \ref{tab:sabdab1}, all of our proposed infill methods now outperform RefineGNN in terms of AAR metric, while reprogBert also provides second best diversity.  
}
\label{tab:sabdab2}
\end{table*}

We included the following baseline methods to compare against our BERT-based infilling models. \emph{LSTM} from \citep{saka2021antibody} and \citep{akbar2022silico}, which, similar to ours, is a sequence-only model, however of smaller capacity, having a single attention layer between the input and output layers. \emph{AR-GNN} - autoregressive graph neural network \citep{jin2021iterative}, which is a sequence and structure-based model, at each step first it predicts the amino acid, followed by the edge generation between the current and all the past residues. \emph{RefineGNN} \citep{jin2021iterative} is a model that designs protein sequence and 3D structure of
CDR together as graphs. At each step the method predicts residues autoregressively and simultaneously refines the predicted global structure, which in turn helps in subsequent residue prediction. To improve computational efficiency, they employ coarse-grained modeling by clustering every predefined number of context residues in a block, thus reducing the size of the computational graph. \emph{AbLang} \citep{Olsen2022}, a language model trained on the antibody sequences in the Observed Antibody Space (OAS) \citep{Kovaltsuk18}, and which was designed to restore missing residues in antibody sequences. 

Additionally, to better evaluate the proposed ReprogBert model, we propose our own baseline approaches for the sequence-based infilling task. The first is in-domain protein model \emph{ProtBert} \citep{Elnaggar2020}, which focuses on task adaptation. It is a specialized protein model that has been
pretrained on millions of protein sequences and therefore is well suited for antibody CDR infilling task. The second is the English language BERT (\emph{EnglishBert}) model, whose out-of-domain language token embeddings are replaced with in-domain amino acid embeddings (see Fig.~\ref{fig:baselines_overview} in Appendix for details), thus the goal here is the domain adaptation. 

We emphasize that among the three proposed infilling approaches, during training all the parameters of the ReprogBert remain frozen except the two linear projection matrices $\theta$ and $\gamma$, which are learned and optimized. On the other hand, all the parameters of ProtBert and EnglishBert, as in typical finetuning,  are still updated and optimized for the CDR infilling task.

\subsection{Structural Antibody Database (SabDab)}
SabDab \citep{gkt1043} is a dataset containing antibody sequences and the corresponding 3D structure information, annotated with several properties like gene details, heavy and light chain pairings, CDR location, etc. For this experiment, we used the dataset curated by \citep{jin2021iterative} and the statistics are shown in Table \ref{tab:sabdab_stats}. The evaluation results are shown in Tables~\ref{tab:sabdab1}, \ref{tab:sabdab2}, and \ref{tab:sabdab3}. We note that the values for PPL and RMSD metrics for LSTM, AR-GNN and Refine-GNN are from the published results \citep{jin2021iterative}. Comparing across the three experiments, we can see that CDR-H1, CDR-H2 and CDR-H3 estimations are progressively harder problems, which is reflected in the drop of AAR across all the methods. Among the proposed infill methods, % overall outperform other baselines, where 
Except LSTM and AR-GNN, all models achieve over 30\% AAR, implying consistency with the ground-truth sequence. %ProtBert achieves the highest AAR across all experiments. We also can see that 
ReprogBert has a good recovery accuracy and at the same time generates very diverse and lowest perplexity CDR sequences, while finetuned Bert models generate less diverse and higher perplexity sequence consistent with earlier-reported performance gap of finetuning. % with one of the lowest perplexities, reflecting their  naturalness. 
We emphasize that the performance of the Bert-based models  is without the access to the available 3D structure information. RefineGNN, on the other hand, using both sequence and structure constraints, overall preforms competitively, generating CDR sequences that are  accurate and diverse. Nevertheless, the advantage of ReprogBert is more prominent for longer CDR-H3, which is the hardest design task of all three, where ReprogBert evidently outperforms RefineGNN in term of perplexity, AAR, and diversity, while maintaining structural integrity. Also observe that AbLang has a consistently lower recovery rate, and even dropping below 30\% threshold, particularly for longer CDR-H3. Such performance might be partially due to a mismatch in the training data of AbLang. 
Finally, in Fig.~\ref{tab:sabdab123} we show the results of all three CDRs infilling at once. The BERT-based models are not architecturally limited to a single CDR generation, in contrast to Refine-GNN, therefore can infill multiple regions at once with similar high recovery, structural consistency,  and diversity scores.

\begin{table*}[!t]
\centering
\resizebox{0.7\textwidth}{!}{%
\begin{tabular}{@{}lcccccccccc@{}}
\toprule
                  & \multicolumn{9}{c}{SabDab CDR-H3}\\
                    \cmidrule(lr){2-10} 
                  & PPL & PPL-ProGen & RMSD & RMSD-AF & RMSD-IF & TM-AF & TM-IF & AAR & DIV\\
\midrule
LSTM             & 9.20  & --    & --   & -- & -- & -- & -- & --  & -- \\
AR-GNN           & 9.44  & --    & 3.63 & -- & -- & -- & -- & -- & --  \\
Refine-GNN       & 8.38  & 7.2 & 2.50 & 5.62 & 3.43 & 85.0 & 94.0  & 28.2  & 25.7  \\
AbLang           & --    & -- & --    & --   & --   & -- & --  & 22.0  & \cellcolor[HTML]{C0C0C0}71.3  \\
ProtBert         & --   & 6.8 & --     & 5.40 & 3.39 & 85.2 & 94.0 & \cellcolor[HTML]{C0C0C0}41.5 & 14.5 \\
EnglishBert      & --   & \cellcolor[HTML]{EFEFEF}5.9 & --    & 5.53 & 3.26 & 84.9 & 94.0 & \cellcolor[HTML]{EFEFEF}35.6 & 59.8 \\
ReprogBert       & --   & \cellcolor[HTML]{C0C0C0}5.4 & --   & 5.54 & 3.44 & 85.1 & 94.0 & 32.6  & \cellcolor[HTML]{EFEFEF}67.4 \\
\bottomrule
\end{tabular}%
}
\caption{Evaluation results on the SabDab dataset for CDR-H3. As compared to CDR-H1 ( Fig.~\ref{tab:sabdab1}) and CDR-H2 (Fig.~\ref{tab:sabdab2}), longer CDR-H3 design  is more  challenging, which shows a drop in AAR across all the methods. ReprogBert clearly outperforms  RefineGNN on this hard task, as evident from lower PPL, better AAR, and better diversity. 
}
\label{tab:sabdab3}
\end{table*}

\begin{figure}[!t]
\centering
\includegraphics[width=0.9\linewidth]{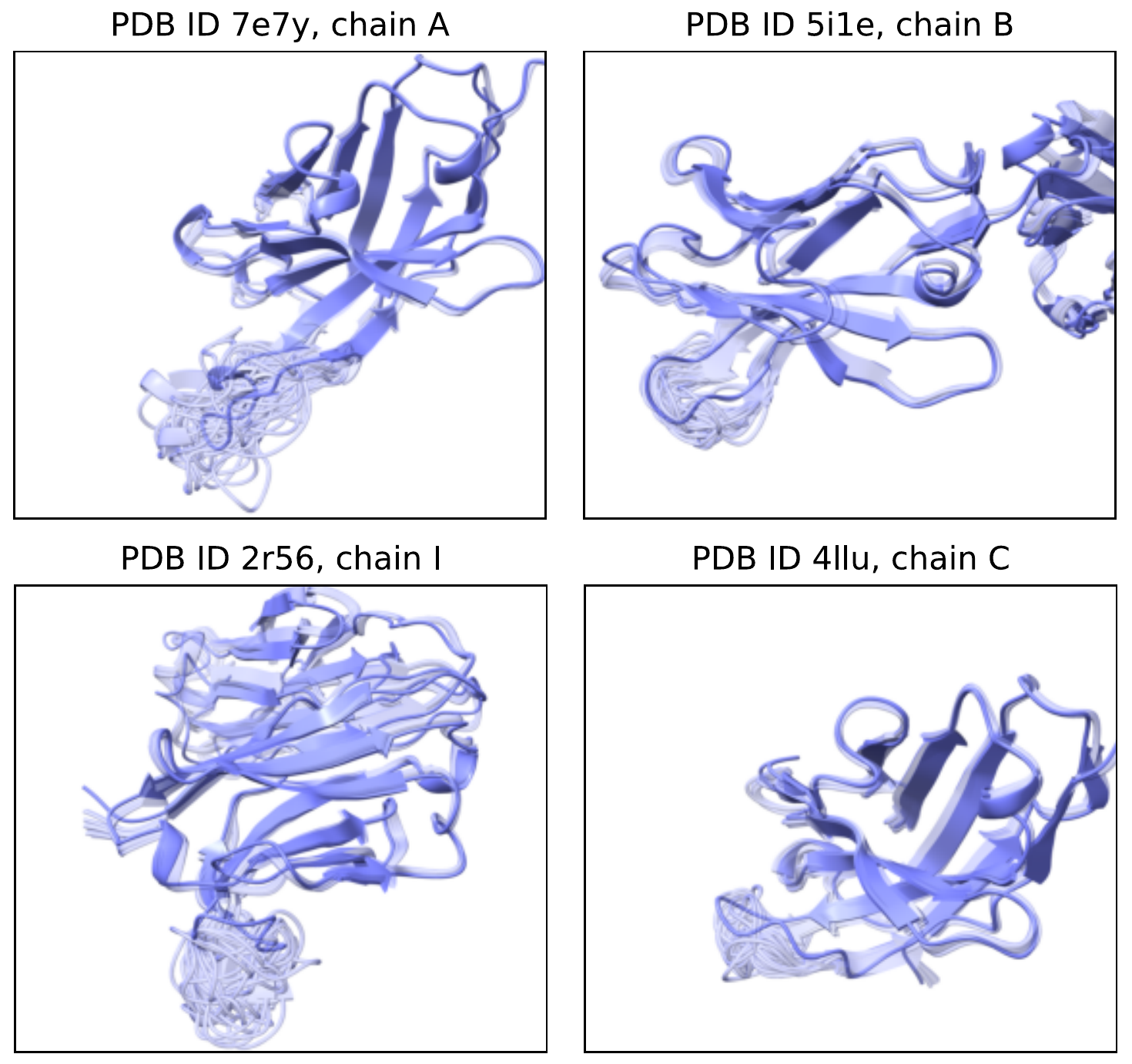}
\caption{AlphaFold-estimated 3D structures of the proteins generated by the ReprogBert model on SabDab dataset. Each plot shows 30 generated CDR samples for a specific PDB ID. % where the CDR-H3 part of the input has been masked and the model then generates CDR-H3 sequence. 
The ground truth and the generated CDR are shown on the bottom part of each figure using solid and faded colors, respectively. As can be seen, CDR-H3 part shows high structural diversity, confirming the same findings as in Table \ref{tab:sabdab3}, i.e., that ReprogBert achieves high recovery rate while maintaining the highest sequence diversity.}
\label{fig:proj_structures}
\end{figure}

Since our BERT-based infill models do not estimate protein structure, we use AlphaFold \citep{jumper2021highly} and IgFold~\citep{ruffolo2022fast} to estimate 3D structure from the generated sequence and compute TM and RMSD scores with respect to groundtruth native structure.  We can see from the Tables~\ref{tab:sabdab1}, \ref{tab:sabdab2}, \ref{tab:sabdab3}, and  \ref{tab:sabdab123} that all the methods have similar structural consistency results (TM and RMSD-AF). However, these values are consistently higher when compared to RMSD for ``natively'' predicted structure (AR-GNN and Refine-GNN), which is likely due to the estimation errors introduced by the AlphaFold or IgFold algorithm. Since RefineGNN focuses on recovering both groundtruth sequence and structure, it does so by  sacrificing  exploration of the broader sequence space accessible to a given structure \citep{tian2017many}, which is not the case for  ReprogBert.  

\begin{figure}[!t]
\centering
\includegraphics[width=0.99\linewidth]{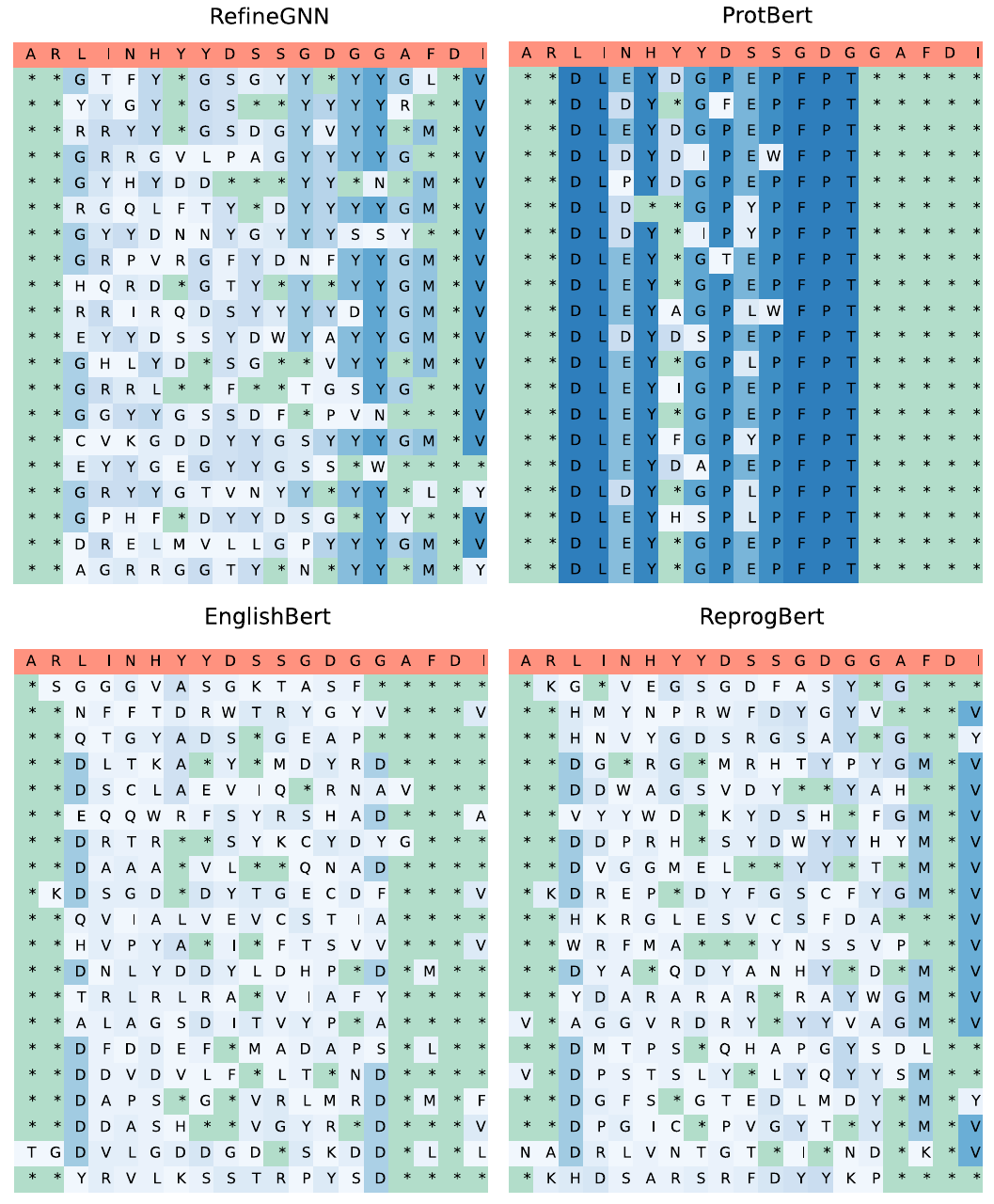}
\caption{Visualization of the sequence recovery and diversity metrics for generated CDR-H3 (PDB ID 7e7y) across different models. The top row (in red) shows the ground truth CDR-H3 , while the following 20 rows correspond to the generated CDR-H3s. % by each of the model. 
The green cell with the star symbol represents the same amino acid as in the ground truth, while the white/blue cell shows new and different generated residues. The darker shade of the blue cell represents the frequency of the amino acid in that column.  }
\label{fig:recdiv_7e7y}
\end{figure}

To further qualitatively illustrate the effect of recovery and diversity on the sampled sequences, we show in Fig.~\ref{fig:proj_structures} AlphaFold-generated 3D structures of the protein sequences generated by the ReprogBert model. High structural diversity of the CDR-H3 is clearly visible by the coverage of the CDR-H3 ensemble (ground-truth shown using opaque while generated shown as transparent). Fig.~\ref{fig:recdiv_7e7y} presents a visualization of sequence similarity (in green)/diversity (in white to blue) across models.
For example, for ProtBert the third column has a residue D in all the rows (high frequency), thus having the darkest shade, while for ReprogBert the last column has only two generated Y's (low frequency), thus colored in the light shade of blue. Therefore, the method with the high recovery and high diversity rates will have many green and light blue cells. Comparing with Table \ref{tab:sabdab3}, we indeed see that ReprogBert has highest diversity represented by the largest number of light blue cells, at the same time ProtBert has most green cells (highest AAR), but also many dark blue cells (low diversity). It can also be seen that RefineGNN has lower diversity and lower recovery, as compared to ReprogBert. Further, the 2D kernel density plot as a function of isoelectric point (pH when net charge is 0) and length of CDR-H3 shown in Figure \ref{fig:isoelectric} in Appendix implies ReprogBert maintains highest physicochemical similarity to the natural CDRs.

\begin{table*}[!ht]
\centering
\resizebox{0.63\textwidth}{!}{%
\begin{tabular}{@{}lcccccccc@{}}
\toprule
                  & \multicolumn{7}{c}{SabDab CDR-H1,2,3}\\
                    \cmidrule(lr){2-8} 
                  & PPL-ProGen & RMSD-AF & RMSD-IF & TM-AF & TM-IF & AAR  & DIV\\
\midrule
AbLang           & --  & --   & -- & -- & --  &  40.1 &  \cellcolor[HTML]{EFEFEF}54.2 \\
ProtBert         & \cellcolor[HTML]{EFEFEF}4.9  & 4.8   & 2.62 & 85.0 & 94.3  &  \cellcolor[HTML]{C0C0C0}57.6 &  8.2 \\
EnglishBert      & 5.2  & 4.83   & 2.62 & 85.0 & 94.2      & \cellcolor[HTML]{EFEFEF}56.3 & 8.3 \\
ReprogBert       & \cellcolor[HTML]{C0C0C0}3.9  & 4.95   & 2.73 & 84.8 & 94.0   & 42.4  & \cellcolor[HTML]{C0C0C0}57.4 \\
\bottomrule
\end{tabular}%
}
\caption{Evaluation results on challenging task of generating the three heavy chain CDR loops  at once using the SabDab dataset. 
The reprogrammed model showed the lowest perplexity, good structural consistency, and the highest sequence variability %among the three proposed methods
.}
\label{tab:sabdab123}
\end{table*}

\subsection{Antigen-Specific Antibody Design}

\begin{table}[!t]
\centering
\resizebox{0.49\textwidth}{!}{%
\begin{tabular}{@{}c|c|cccc@{}}
\toprule
Dataset   & CDR    & Train & Validation & Test & \multicolumn{1}{l}{Average CDR length} \\ \midrule
RabD      & CDR-H3 & 8646  & 98         & 58   & 14.5                                   \\
CoV-AbDab & CDR-H3 & 2282  & 291        & 291  & 15.7                                   \\ \bottomrule
\end{tabular}%
}
\caption{Statistics of Rosetta Antibody Design (RabD) and Coronavirus Antibody Database (CoV-AbDab) datasets for CDR-H3.}
\label{tab:rabd_cov_stats}
\end{table}

The goal here is to design a CDR that binds a given antigen, given the antibody sequence template. For this experiment, we used the dataset curated by \citep{jin2021iterative}, statistics of which is shown in Table \ref{tab:rabd_cov_stats}. In particular it consists of all the SabDab \ref{tab:rabd_cov_stats} for training, excluding sequences in the same cluster as test antibodies, which were proposed by \citep{adolf2018rosettaantibodydesign}. In addition to the earlier defined baselines, for this experiment, similar to \citep{jin2021iterative}, we compared against a physics-based baseline, RabD~\citep{adolf2018rosettaantibodydesign}, which first grafts a CDR from an internal database into the groundtruth antibody structure, followed by iterations of amino acid substitutions and energy minimization.  
The results are shown in Table~\ref{tab:rabd}. The values for PPL, RMSD, and AAR metrics for RabD, LSTM, AR-GNN and Refine-GNN baselines are from \citep{jin2021iterative}. 
ReprogBert shows the best diversity rate with accurate sequence recovery and structural consistency.

\begin{table*}[!ht]
\centering
\resizebox{0.75\textwidth}{!}{%
\begin{tabular}{@{}lcccccccccc@{}}
\toprule
                  & \multicolumn{9}{c}{RabD CDR-H3}\\
                    \cmidrule(lr){2-10} 
                  & PPL & PPL-ProGen & RMSD & RMSD-AF & RMSD-IF & TM-AF & TM-IF & AAR & DIV\\
\midrule
RabD             & 9.20  & --     & --   & -- & -- & -- & -- &  28.53    & -- \\
LSTM             & 9.20  & --     & --   & -- & -- & -- & -- &  22.53     & -- \\
AR-GNN           & 9.44  & --     & 3.63 & -- & -- & -- & -- &  23.86     & --  \\
AbLang           & --  & --  & -- & -- & -- & -- & -- &  21.3     & \cellcolor[HTML]{C0C0C0}70.9  \\
Refine-GNN       & 8.38  & \cellcolor[HTML]{C0C0C0}4.7  & 2.50 & 5.06 & 2.52 & 82.9 & 96.0 &  35.4     & 31.1  \\
ProtBert         & --    & 7.7  & --   & 5.42 & 2.35 & 82.3 & 96.2   & \cellcolor[HTML]{EFEFEF}53.1   & 11.6 \\
EnglishBert      & --    & 7.8  & --   & 5.34 & 2.19 & 82.4 & 96.3 & \cellcolor[HTML]{C0C0C0}54.9      & 10.1 \\
ReprogBert       & --    & \cellcolor[HTML]{EFEFEF}5.1  & --   & 4.72 & 2.47 & 83.0 & 96.1   & 36.3  & \cellcolor[HTML]{EFEFEF}62.1 \\
\bottomrule
\end{tabular}%
}
\caption{Evaluation results on the RabD dataset for CDR-H3. The ReprogBert achieves the best diversity rate with accurate sequence recovery and structural consistency. 
}
\label{tab:rabd}
\end{table*}

\begin{table*}[!t]
\centering
\resizebox{0.6\textwidth}{!}{%
\begin{tabular}{@{}cccc|ccc@{}}
\toprule
\multirow{2}{*}{} & \multicolumn{3}{c|}{Training on CoV-AbDab} & \multicolumn{3}{c}{ Training on CoV-AbDab + SabDab} \\ \cmidrule(l){2-7} 
                  & PPL-ProGen   & AAR    & DIV   & PPL-ProGen      & AAR     & DIV      \\ \midrule
ProtBert          & 6.0           & \cellcolor[HTML]{C0C0C0}50.7   & \cellcolor[HTML]{EFEFEF}13.6  & 7.8             & \cellcolor[HTML]{C0C0C0}49.6    & 10.7     \\
EnglishBert       & 6.3            & \cellcolor[HTML]{EFEFEF}49.3  & 9.5   & 8.2              & \cellcolor[HTML]{EFEFEF}49.2     & \cellcolor[HTML]{EFEFEF}11.0     \\
ReprogBert        & 5.7           & 39.3 & \cellcolor[HTML]{C0C0C0}60.2  & 4.9              & 37.3     & \cellcolor[HTML]{C0C0C0}64.1     \\ \bottomrule
\end{tabular}%
}
\caption{Evaluation results on the CoV-AbDab dataset for generated CDR-H3. Since no ground truth structure is available for this dataset, the structure consistency metrics are not computed.}
\label{tab:cov}
\end{table*}

\subsection{Coronavirus Antibody Database (CoV-AbDab)}

We also show generality of our reprogramming approach on CoV-AbDab \citep{Raybould2021}, a public database documenting all published and patented antibodies and nanobodies able to bind to coronaviruses, including SARS-CoV2 and SARS-CoV1. We used the dataset curated by \citep{jin2021iterative} (see Table \ref{tab:rabd_cov_stats}).
The evaluation results are shown in Table~\ref{tab:cov}, where only the sequence-based metrics are presented since the ground truth structure information is unavailable for this task. The results are presented for the case of training only on CoV-AbDab and the case of training on both CoV-AbDab and SabDab datasets, showing overall similar trend, i.e. %with ProtBert being the most accurate in AAR evaluation, while
ReprotBert achieving the highest diversity while maintaining good sequence recovery and low perplexity.

The second step of our evaluation is to measure the ability of the generated antibodies to neutralize SARS-CoV2 virus, for which we follow the setup of \citep{jin2021iterative}. Specifically, we  employ  the  neutralization classifier, composed of SRU encoder \citep{lei2021attention}, pooling and feed-forward network, as provided in \citep{github_refinegnn}, together with the iterative target augmentation (ITA) framework \citep{yang2020improving}. The goal is to additionally fine-tune the infilling models to generate CDRs resulting into better neutralizing  antibodies, as measured by the classifier. Table \ref{tab:neutr} presents the results. Note that the performance values for the neutralization classifier, LSTM, AR-GNN and Refine-GNN are from  \citep{jin2021iterative}, for which they pretrained these models on SabDab dataset followed by the training on CoV-AbDab.  As can be seen from the table, under both training scenarios, our ReprogBert infilling method gets the largest improvement over the original neutralization classifier, achieving 75.6 \% and 76.7 \% neutralization scores, respectively.

\begin{table}[!t]
\centering
\resizebox{0.35\textwidth}{!}{%
\begin{tabular}{@{}ccc@{}}
\toprule
            & \multicolumn{2}{c}{Neutralization Score} \\ \midrule
Model       & CoV-AbDab    & CoV-AbDab + SabDab    \\ \midrule
Original    & --          & 69.3                 \\
LSTM        & --          & 72.0                 \\
AR-GNN      & --          & 70.4                 \\
Refine-GNN  & --          & \cellcolor[HTML]{EFEFEF}75.2                 \\
ProtBert    & \cellcolor[HTML]{EFEFEF}72.7        & 74.7                 \\
EnglishBert & 70.5        & 71.0                 \\
ReprogBert  & \cellcolor[HTML]{C0C0C0}75.6        & \cellcolor[HTML]{C0C0C0}76.7                 \\ \bottomrule
\end{tabular}%
}
\caption{Neutralization of SARS-CoV-2 virus as predicted by the pre-trained SARS-CoV-1 / SARS-CoV-2 classifier. The neutralization score is defined as the predicted probability of a given antibody to neutralize the SARS-CoV-2 virus, as measured by the neutralization classifier.}
\label{tab:neutr}
\end{table}

\section{Limitations of our work}
In this section, we discuss the challenges faced by our proposed model in handling larger protein infilling tasks, its dependence on pre-trained language models, and the potential limitations stemming from a restricted training dataset which may impact the experimental validation of generated antibody sequences in wet labs.
\begin{itemize}
    \item Limited performance on larger protein infilling tasks: While the proposed model shows promising results on smaller CDR regions, it may not perform well on larger protein infilling tasks, when the context becomes too small, and where more complex structural dependencies need to be considered. However, the better performance on longer CDR-H3 of ReprogBert, compared to baselines, is noteworthy and promising for extending it to loop design task in general. One way to handle that would be to force attention close to the region to be infilled, which will be future work.
    \item Dependence on pre-trained language models: The proposed model relies on existing pre-trained language models, which may not capture all the domain-specific features required for protein infilling tasks. Reprogramming provides a framework to learn mapping between amino acids and English vocabularies, given that the protein sequences capture structure and functional information \citep{Rives21} and, protein sequences share linguistic features such as Zipfian distribution and existence of grammar \citep{Yu19}.
    \item Since the model was trained on a limited set of available antibody data, it may not capture all the complex interactions and structural constraints required for functional antibodies. The generated antibody sequences therefore may fail experimental validation of wet lab.
\end{itemize}

\section{Conclusion}
In this work we introduced ReprogBert, a reprogramming framework
leveraging pretrained English language models for protein sequence
infilling.  Specifically, we formulated variable CDR loop design  as a template-infilling, where the
template is provided by the constant region of the antibody.  
Results show promising performance, when compared to existing sequence and graph-based deep generative baselines, as well as straightforward task-specific and domain-specific finetuned pre-trained language models.  Over multiple benchmarks, our ReprogBert model upholds structural integrity, sequence recovery, and naturalness, while achieving high novelty and diversity of the generated sequences. The improvement is more obvious for the longer CDR-H3. ReprogBert can also handle multiple CDR infilling at once with losing performance. The generated sequences  demonstrate enhanced antigen binding specificity and virus neutralization ability in silico. Analysis of the amino acid embeddings learned by ReprogBert from English token embeddings reveals efficient mapping between the two domains as well as naturally occurring clustering of amino acids with meaningful biological properties.
Finally, it is worth emphasizing the high data-efficiency of the reprogrammed model, which results from having only a few training parameters (consisting of two linear projection matrices) that can be efficiently trained in the data-scarce domains, such as antibody design, while still leveraging information from large out-of-domain language pretraining. This advantage allows the reprogrammed  English language model to efficiently and effectively adapt to the antibody sequences, and perform competitively or better with respect to other supervised or finetuned baselines that either learn from both sequences and structures, or requires more expensive finetuning, or show performance degradation on chanllening design tasks.

\bibliography{references}
\bibliographystyle{icml2023}

\clearpage
\appendix
\onecolumn

\section{Related Work on Protein Design}
Protein design involves the design of new protein sequences that fold to a desired 3D structure and/or exhibit a specific function. Computational techniques for designing novel and diverse proteins are an active area of research. Physics based methods that rely on energy minimization have been proposed for designing general proteins ~\citep{leaverfayrosetta, huang2011rosettaremodel}, as well as specifically for 
 antibodies~\citep{pantazes2010optcdr,li2014optmaven, adolf2018rosettaantibodydesign}, but these are computationally expensive. Recently, generative deep learning techniques like Generative Adversarial Networks~\citep{goodfellow2020generative}, Variational Autoencoders~\citep{kingma2013auto}, Graph Neural Networks~\citep{scarselli2008graph, gilmer2017neural}, autoregressive language models (LSTM and Transformer based)~\citep{vaswani2017attention}, and diffusion based models~\citep{ho2020denoising} have been used for protein and antibody design~\citep{wang2018computational,akbar2022silico,amimeur2020designingantibody,eguchiigantibody,shin2021proteinantibody,  kong2022conditional,fu2022antibody,syrlybaeva2022deep, lee2022proteinsgm, anand2022protein}. Some representative works  are discussed below. \citep{ingraham2019generative} and \citep{cao2021fold2seq} proposed a graph and a multimodal transformer based model, 
 respectively, for designing proteins conditioned on the backbone structure/fold. ~\citep{karimi2020novo}, developed a guided conditional Wasserstein Generative Adversarial Networks (gcWGAN) for fold based protein design. Another method that uses GANs to generate a distance matrix representation of proteins from which 3D coordinates can be recovered was proposed by ~\citep{anand2018generative}. Variational autoencoder based methods have also been proposed for conditional generation of protein sequences~\citep{greener2018design, das2021accelerated} and for direct generation of 3D coordinates of immunoglobulin proteins~\citep{eguchiigantibody}. 
 
Several of the above-mentioned architectures have been extended to the specific problem of antibody design, which is considered challenging due to focus on designing long, variable, and unstructured CDRs. \citep{melnyk2021benchmarking} provides  benchmarking of several deep generative models on antibody design. Recently, ~\citep{jin2021iterative} proposed an iterative refinement graph neural network for jointly designing the sequence and 3D structure of the CDR regions of antibodies for improving its properties. A  deep generative model that jointly models sequences and structures of CDRs based on diffusion processes and equivariant neural networks has been proposed in \citep{Luo2022}. A geometry-constrained energy-based model has been suggested by ~\citep{fu2022antibody}.

Other approaches for protein design include modeling it as a constraint satisfaction problem~\citep{strokach2020fast}, equivariant 3D translation~\citep{kong2022conditional} and by using combinatorial bayesian optimization~\citep{khan2022antbo}. 
%In this work, we seek to design antibodies by infilling the CDR portion of the sequence using model reprogramming on antibody sequences.

\section{Overview of Proposed Baseline Models}

\begin{figure}[!ht]
\centering
\includegraphics[width=0.99\linewidth]{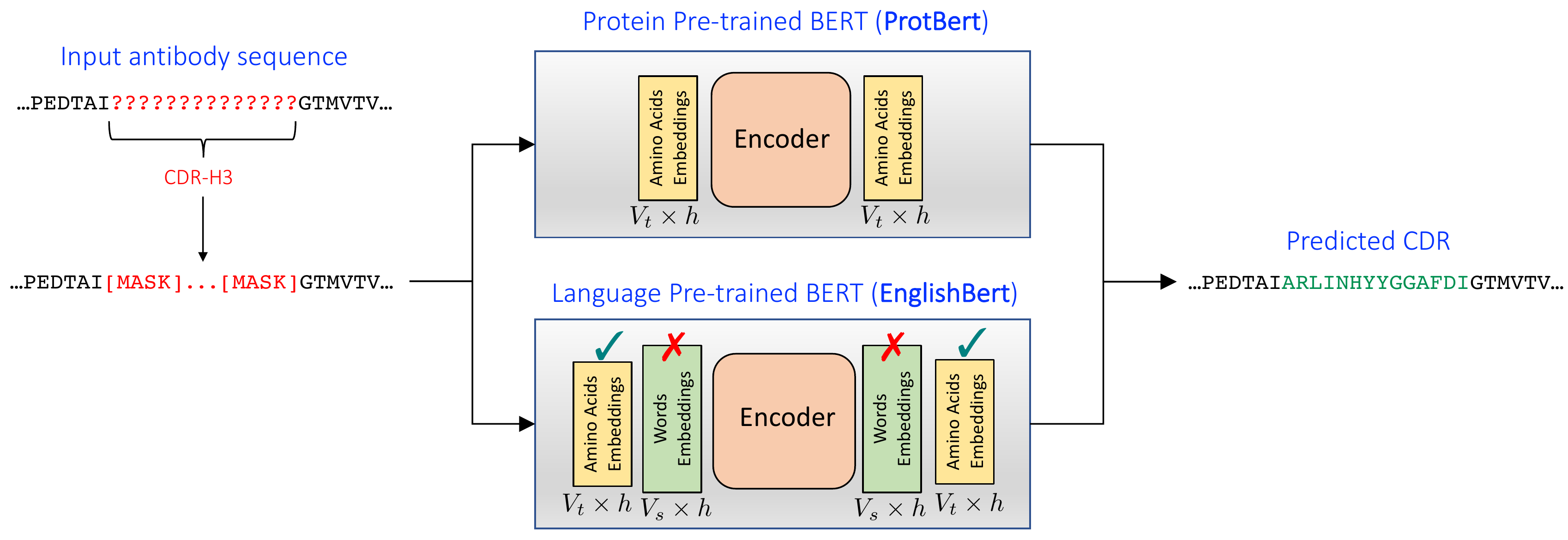}
\caption{Baseline methods proposed for protein sequence infilling. Given an input antibody sequence, where part of the amino acids is missing (e.g., CDR-H3), the goal is to infill them using information from the rest of the protein. The infilling problem is formulated similar to the masked-language modeling task, where the missing amino acids are marked with a token $\mathtt{\langle MASK \rangle}$ and the model generates amino acids token to infill them. These are sequence-only methods and do not rely on any structure information during generation process. The top diagram shows \emph{ProtBert}, the BERT model that has been pretrained on the protein sequences and therefore can be applied to the protein infilling task as is (the entire model is still fine-tuned on the downstream infilling task). The bottom diagram shows traditional English language BERT model (\emph{EnglishBert}), whose incompatible word embeddings ($V_s \times h$, $V_s$ is the number of language tokens, $h$ - latent model dimension) are swapped with the trainable amino acid embeddings ($V_t \times h$, $V_t$ is the number of amino acid tokens). The full model is then fine-tuned on the infilling dataset.}
\label{fig:baselines_overview}
\end{figure}

Figure \ref{fig:baselines_overview} shows diagrams of the proposed baseline BERT-based infilling models: ProtBert, a specialized model that has been pretrained on millions of protein sequences and EnglishBert, the traditional English language model, where we replaced word embeddings with new learnable amino acid embeddings. Similar as our main proposed method, ReprogBert, these two models are sequence-only methods and they use maskings to infill the regions of interest.

\FloatBarrier

\section{Model Architecture and Training}
In Table~\ref{tab:model_info} we present the architectural details of our BERT-based models for the protein sequence infilling, while Table~\ref{tab:training} shows the settings used for model training.

\begin{table}[!ht]
\centering
\resizebox{0.8\textwidth}{!}{%
\begin{tabular}{@{}cccccclc@{}}
\toprule
Model &
  \begin{tabular}[c]{@{}c@{}}Number of \\ parameters\end{tabular} &
  \begin{tabular}[c]{@{}c@{}}Number of\\  layers\end{tabular} &
  \begin{tabular}[c]{@{}c@{}}Hidden \\ layer size\end{tabular} &
  \begin{tabular}[c]{@{}c@{}}Number of\\  heads\end{tabular} &
  \begin{tabular}[c]{@{}c@{}}Vocab\\ size\end{tabular} &
  \multicolumn{1}{c}{Pretraining Data} &
  Reference \\ \midrule
ProtBert &
  420M &
  30 &
  1024 &
  16 &
  30 &
  \begin{tabular}[c]{@{}l@{}}BFD100 \\ (572 GB, 2 bil proteins)\\ \\ Uniref100\\ (150 GB, 216 mil proteins)\end{tabular} &
   \citep{devlin2018bert} \\ \midrule
\begin{tabular}[c]{@{}c@{}}EnglishBert / ReprogBert\\ (based on HF bert-base-uncased)\end{tabular} &
  110M &
  12 &
  768 &
  12 &
  \begin{tabular}[c]{@{}c@{}}30522\\ (english)\\ \\ 30\\ (protein)\end{tabular} &
  \begin{tabular}[c]{@{}l@{}}English Wikipedia\\ (40 GB, 6.5 mil sentences)\\ \\ BookCorpus \\ (6 GB, 74 mil sentences)\end{tabular} &
   \citep{Elnaggar2020} \\ \bottomrule
\end{tabular}%
}
\caption{Architectural details of the BERT-based model for protein sequence infilling. Note that for ReprogBert the number of trainable parameters is defined by the two $\mathbb{R}^{30522\times 30}$ matrices.}
\label{tab:model_info}
\end{table}

\begin{table}[!ht]
\centering
\resizebox{0.3\textwidth}{!}{%
\begin{tabular}{@{}ccc@{}}
\toprule
Learning rate & Batch size & Optimizer \\ \midrule
$1e^{-5}$          & 32         & Adam      \\ \bottomrule
\end{tabular}%
}
\caption{Training details for ProtBert, EnglishBert and ReprogBert. For example, for SabDab dataset to reach the best performance it took 5 hours for ReprogBert, 6 hours for EnglishBert and 14 hours for ProtBert, which is equivalent to approximately 1800 epochs (134 minibatch iterations per epoch). We trained all models on a single A100 40GB GPU. Average inference time per protein sequence is 0.02 seconds for ProtBert, and 0.008 seconds for ReprogBert and EnglishBert (as measured on the test set of SabDab for CDR-H3 infilling). For reference, the average inference time for RefineGNN is 0.004 seconds, which is comparable to our ReprogBert.}
\label{tab:training}
\end{table}

\FloatBarrier
\section{Ablation on Data}

In Table ~\ref{tab:abl} we show an ablation results on the effect of training data size on model performance.

\begin{table}[!ht]
\centering
\resizebox{0.5\textwidth}{!}{%
\begin{tabular}{@{}ccccc@{}}
\toprule
\multicolumn{5}{c}{SabDab-H3}                                                    \\ \midrule
\multicolumn{1}{l}{}         & Training data fraction & PPL-ProGen & AAR  & DIV  \\ \midrule
\multirow{5}{*}{ProtBert}    & 1.0                    & 6.8        & 41.5 & 14.5 \\
                             & 0.8                    & 6.7        & 41.3 & 13.1 \\
                             & 0.6                    & 6.6        & 40.9 & 15.9 \\
                             & 0.4                    & 6.4        & 40.5 & 18.9 \\
                             & 0.2                    & 6.6        & 40.3 & 18.4 \\ \midrule
\multirow{5}{*}{EnglishBert} & 1.0                    & 5.9        & 35.9 & 59.8 \\
                             & 0.8                    & 5.9        & 35.1 & 57.9 \\
                             & 0.6                    & 6.5        & 34.2 & 59.6 \\
                             & 0.4                    & 6.4        & 33.6 & 61.4 \\
                             & 0.2                    & 6.5       & 33.1 & 63.5 \\ \midrule
\multirow{5}{*}{ReprogBert}  & 1.0                    & 6.0        & 32.6 & 67.4 \\
                             & 0.8                    & 5.9        & 32.1 & 67.6 \\
                             & 0.6                    & 6.1        & 31.6 & 68.2 \\
                             & 0.4                    & 6.3        & 30.8 & 69.5 \\
                             & 0.2                    & 6.5        & 29.9 & 70.7 \\ \bottomrule
\end{tabular}%
}
\caption{Ablation results on the effect of training data size on model performance. The fractions 1.0, 0.8, 0.6, 0.4 and 0.2 representing progressively smaller subsets of the original SabDab training dataset. It can be seen that as the size of training data drops, the recovery rate also decreases, while the diversity increases (this is expected as now the generated sequences are less accurate). However, for ProtBert, the decrease is slower, likely due to this model being pretrained on large protein dataset, thus retaining its prediction capacity. }
\label{tab:abl}
\end{table}

\FloatBarrier
\section{Ablation on Models}

In Tables ~\ref{tab:abl_models_cdr1}, \ref{tab:abl_models_cdr2}, and \ref{tab:abl_models_cdr3}  we show ablation results on the effect of model sizes and pre-training on the performance of EnglishBert and ReprogBert models when trained and tested on SabDab dataset. One observation we can make is that the larger pre-trained model (bert-large-uncased, 340M) results in lower AAR and higher DIV as compared to using smaller pre-trained bert-base-uncased (110M). This trend is more pronounced for EnglishBert, which sees more abrupt drops in AAR and significant increase in DIV for CDR-H1 and CDR-H2. This is expected as the model becomes less accurate and more random in generating the CDRs. On the other hand, ReprogBert is more stable and we see a smaller change in recovery and diversity. For CDR-H3 both EnglishBert and ReprogBert have similar drop in accuracy accompanied with smaller increase in diversity. We can conclude that increasing the model capacity did not improve performance, likely due to the limited size of available training data. On the other hand, using pre-trained language models is still beneficial as compared to starting from scratch (third column in the Tables). 

\begin{table}[]
\centering
\resizebox{0.7\textwidth}{!}{%
\begin{tabular}{@{}cccccccccc@{}}
\toprule
                                 & \multicolumn{9}{c}{SabDab CDR-H1}                                                                                    \\ \midrule
\multicolumn{1}{c|}{}            & \multicolumn{3}{c|}{Base (110M)}       & \multicolumn{3}{c|}{Large (340M)}      & \multicolumn{3}{c}{Scratch (110M)} \\ \cmidrule(l){2-10} 
\multicolumn{1}{c|}{} & PPL-ProGen & AAR & \multicolumn{1}{c|}{DIV} & PPL-ProGen & AAR & \multicolumn{1}{c|}{DIV} & PPL-ProGen & AAR & DIV \\ \cmidrule(l){2-10} 
\multicolumn{1}{c|}{EnglishBert} & 3.7 & 63.6 & \multicolumn{1}{c|}{5.8}  & 7.5 & 43.8 & \multicolumn{1}{c|}{63.4} & 15.1       & 38.3      & 71.9      \\
\multicolumn{1}{c|}{ReprogBert}  & 3.3 & 56.0 & \multicolumn{1}{c|}{29.1} & 5.2 & 51.5 & \multicolumn{1}{c|}{39.4} & 4.5        & 48.7      & 45.5      \\ \bottomrule
\end{tabular}%
}
\caption{Ablation results on the effect of the model size and pre-training on the performance of EnglishBert and ReprogBert models for CDR-H1.}
\label{tab:abl_models_cdr1}
\end{table}

\begin{table}[]
\centering
\resizebox{0.7\textwidth}{!}{%
\begin{tabular}{@{}cccccccccc@{}}
\toprule
                                 & \multicolumn{9}{c}{SabDab CDR-H2}                                                                                    \\ \midrule
\multicolumn{1}{c|}{}            & \multicolumn{3}{c|}{Base (110M)}       & \multicolumn{3}{c|}{Large (340M)}      & \multicolumn{3}{c}{Scratch (110M)} \\ \cmidrule(l){2-10} 
\multicolumn{1}{c|}{} & PPL-ProGen & AAR & \multicolumn{1}{c|}{DIV} & PPL-ProGen & AAR & \multicolumn{1}{c|}{DIV} & PPL-ProGen & AAR & DIV \\ \cmidrule(l){2-10} 
\multicolumn{1}{c|}{EnglishBert} & 4.0 & 59.1 & \multicolumn{1}{c|}{7.7}  & 8.9 & 40.5 & \multicolumn{1}{c|}{68.7} & 10.3       & 38.4      & 72.7      \\
\multicolumn{1}{c|}{ReprogBert}  & 3.9 & 53.0 & \multicolumn{1}{c|}{37.9} & 8.5 & 50.9 & \multicolumn{1}{c|}{43.4} & 13.8       & 43.6      & 62.8      \\ \bottomrule
\end{tabular}%
}
\caption{Ablation results on the effect of the model size and pre-training on the performance of EnglishBert and ReprogBert models for CDR-H2.}
\label{tab:abl_models_cdr2}
\end{table}

\begin{table}[]
\centering
\resizebox{0.7\textwidth}{!}{%
\begin{tabular}{@{}cccccccccc@{}}
\toprule
                                 & \multicolumn{9}{c}{SabDab CDR-H3}                                                                                    \\ \midrule
\multicolumn{1}{c|}{}            & \multicolumn{3}{c|}{Base (110M)}       & \multicolumn{3}{c|}{Large (340M)}      & \multicolumn{3}{c}{Scratch (110M)} \\ \cmidrule(l){2-10} 
\multicolumn{1}{c|}{} & PPL-ProGen & AAR & \multicolumn{1}{c|}{DIV} & PPL-ProGen & AAR & \multicolumn{1}{c|}{DIV} & PPL-ProGen & AAR & DIV \\ \cmidrule(l){2-10} 
\multicolumn{1}{c|}{EnglishBert} & 5.9 & 35.6 & \multicolumn{1}{c|}{59.8} & 5.5 & 32.2 & \multicolumn{1}{c|}{68.2} & 6.9       & 20.1       & 85.1      \\
\multicolumn{1}{c|}{ReprogBert}  & 5.4 & 32.6 & \multicolumn{1}{c|}{67.4} & 5.6 & 29.7 & \multicolumn{1}{c|}{70.9} & 7.0       & 20.3       & 84.8      \\ \bottomrule
\end{tabular}%
}
\caption{Ablation results on the effect of the model size and pre-training on the performance of EnglishBert and ReprogBert models for CDR-H3.}
\label{tab:abl_models_cdr3}
\end{table}

\FloatBarrier
\section{Antibody Developability Prediction Task}

Here we present additional experimental results on antibody developability. We use the web server from (https://opig.stats.ox.ac.uk/webapps/newsabdab/sabpred/tap) for developability prediction for 35 randomly selected sequences from our ReprogBert-based generations on the SabDab database. Only 2 of them (5.7\%) show red flag for at least one of the five developability metrics, indicating a previously unobserved value for that property, as described in \cite{Raybould19}. This analysis indicates that the generated antibody sequences by ReprogBert do not pose any significant developability concern.

\FloatBarrier
\section{Protein-Protein Docking}

\begin{figure}[!ht]
\centering
\includegraphics[width=0.3\linewidth]{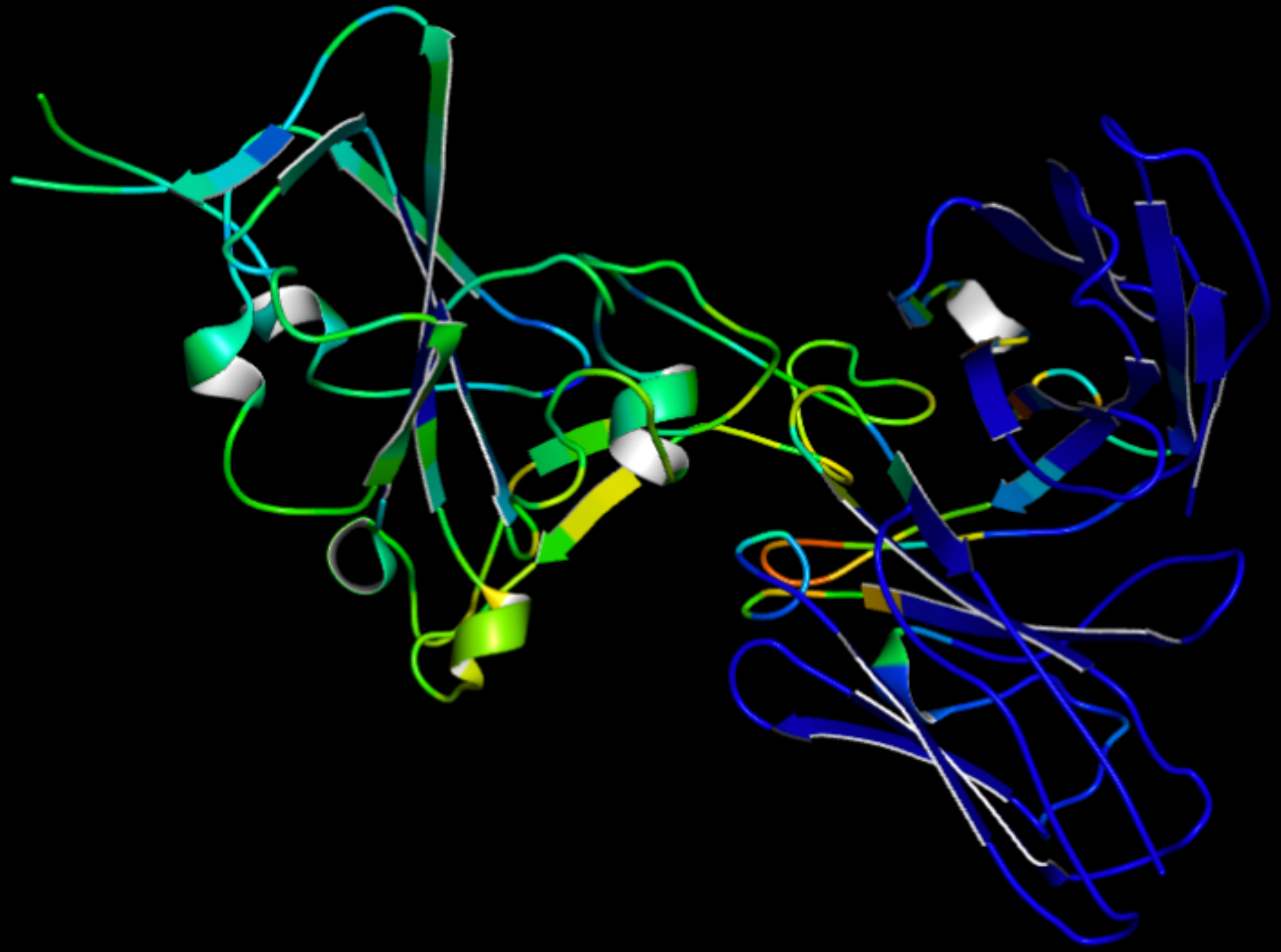}
\caption{Docked structure of BD55-6478 with Spike receptor binding domain (from pdb id: 7l7d)}
\label{fig:ppdocking1}
\end{figure}

\begin{figure}[!ht]
\centering
\includegraphics[width=0.3\linewidth]{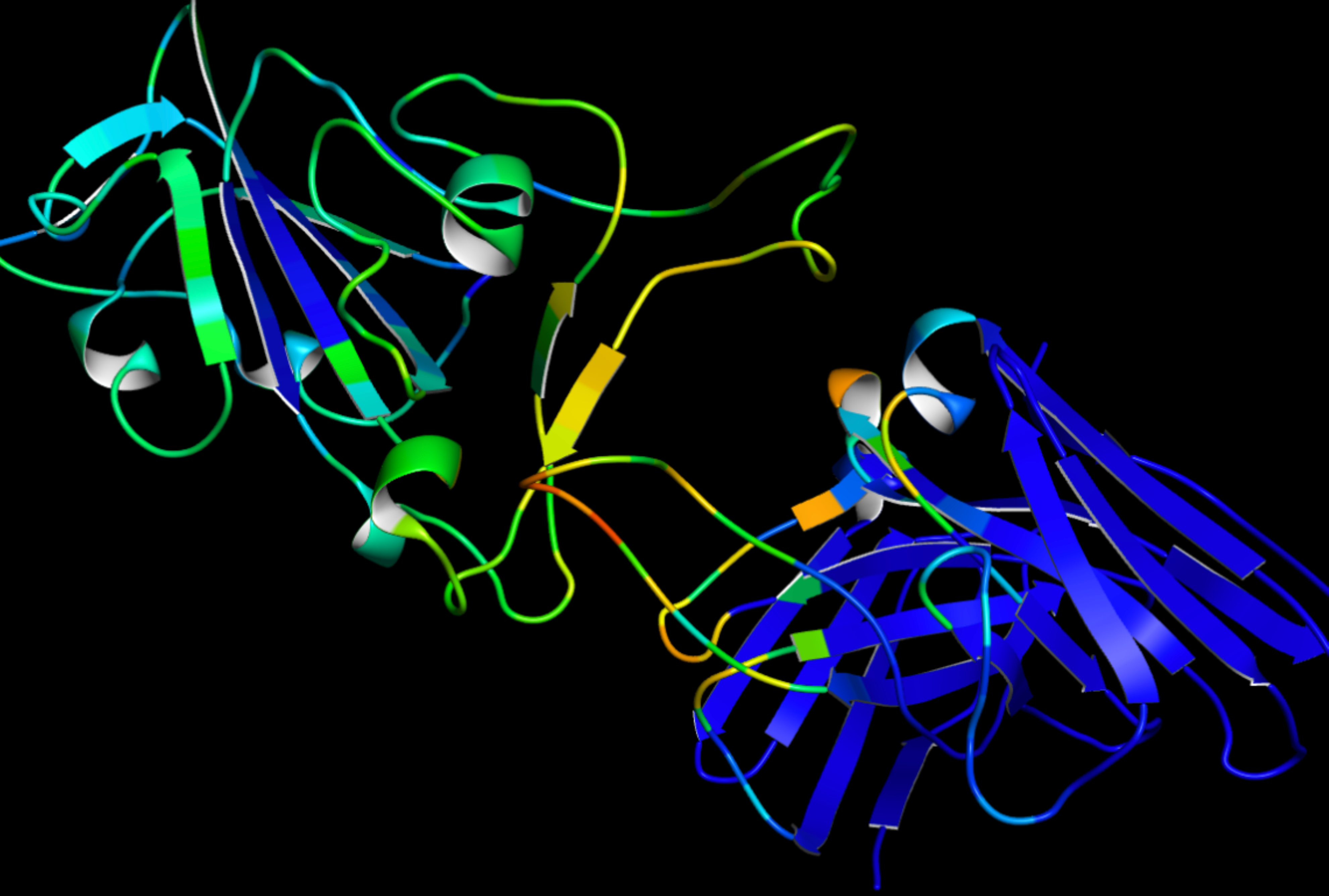}
\caption{Docked structure of BD55-6478 with Spike receptor binding domain (from pdb id: 7l7d)}
\label{fig:ppdocking2}
\end{figure}

For this task we have predicted the structure of the ReprogBert-designed antibody sequence for COVID use-case with Spike receptor binding domain from PDB id: 7l7d by using AbAdapt web server (https://sysimm.org/abadapt/). Two such docked structures are shown as examples, which represent the cluster ($size > 1$ structure) with the best docking score. The antibody residues are colored according to the epitope probabilities using an RGB (high to low) scale. Fig. \ref{fig:ppdocking1} shows a socked structure of BD55-6478 with Spike receptor binding domain (from pdb id: 7l7d), and Fig. \ref{fig:ppdocking2} shows a docked structure of BD56-124 with Spike receptor binding domain (from pdb id: 7l7d).

\FloatBarrier
\section{Examining Amino Acid Embeddings and Mappings}
\label{sec:mapping}
In this section we examine and visualize the embeddings of amino acids and see if there are any naturally forming clustering present. We also examine the mappings between the protein space of amino acids and the space of English tokens of our ReprogBert model. For this, we first review the linear projection method and examine the corresponding embeddings, then we show that it is not easy to use it to visualize the mappings, and propose the alternative based on cross-attention mechanism.
\subsection{Projection}
\label{sec:apendix_proj}

Recall from Section \ref{sec:infill} that we reprogram EnglishBert model by introducing two linear projection matrices 
\begin{align}
\theta &\in \mathbb{R}^{|V_t|\times |V_s|}\\
\gamma &\in \mathbb{R}^{|V_s|\times |V_t|}
\end{align}
to project the target protein domain $x_t$ to source English token domain $x_s$, and then similarly reverse the mapping of the output, we get, respectively:
\begin{align}
x_s &= x_t\theta \\
y_t &= \gamma y_s.
\end{align}

In particular, focusing on the input projection matrix $\theta$ and treating $x_t$ as a one-hot sequence representation of the amino acids of length $N$, i.e., $x_t \in \mathbb{R}^{N\times |V_t|}$, then the sequence representation in the English token domain becomes $x_s \in \mathbb{R}^{N\times |V_s|}$. Then, this representation is projected onto the embedding matrix of the English Bert model $E \in \mathbb{R}^{|V_s|\times d}$ ($d$ is the hidden dimension of English Bert):
\begin{align}
x_s^{E} = x_s E,
\label{eq:emb}
\end{align}
and continue the usual processing through the transformer layers and blocks. 

\begin{figure}[!ht]
\centering
\includegraphics[width=0.99\linewidth]{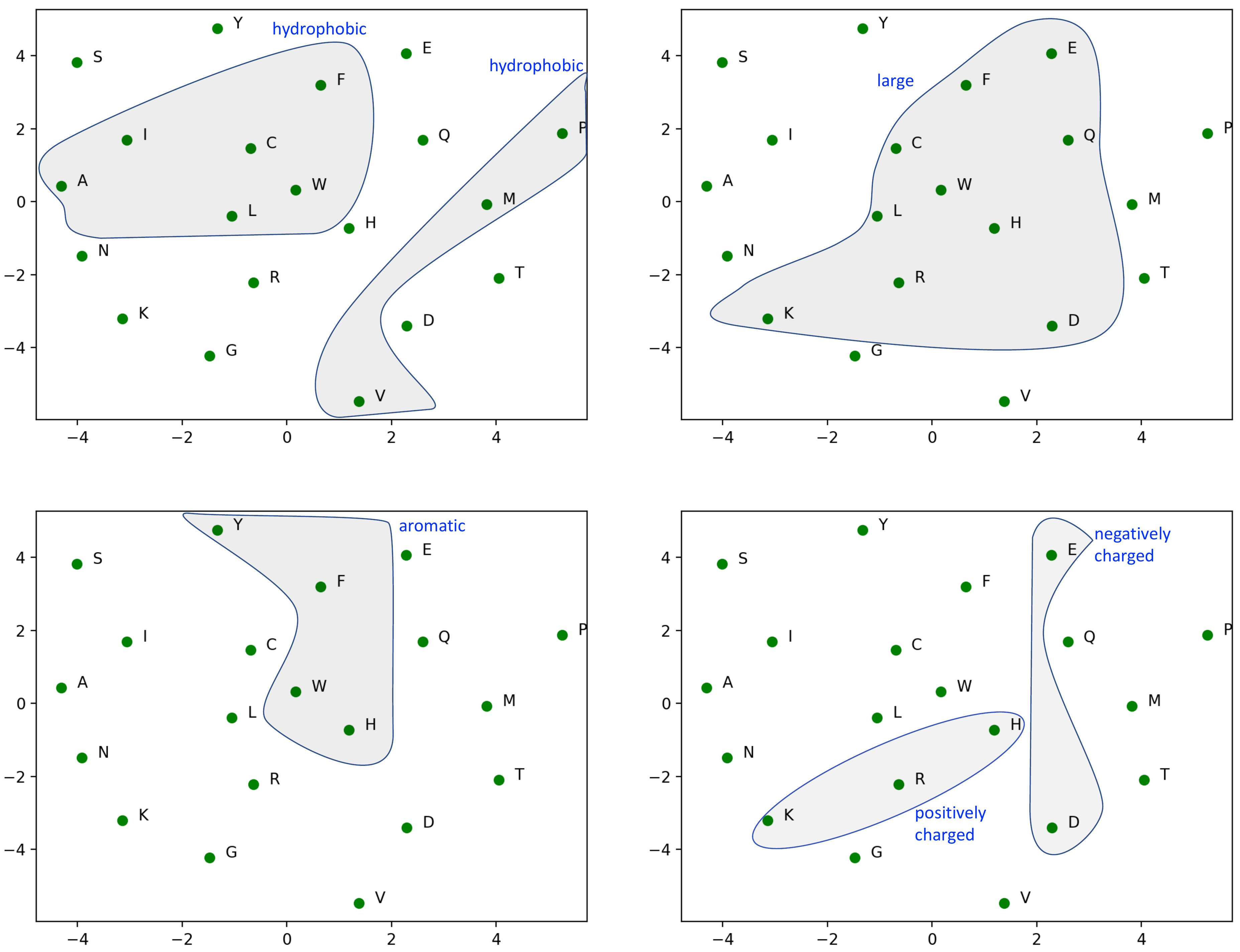}
\caption{Projection of 20 amino acids, defined by embedding matrix $E_{aa}$ (see equation \eqref{eq:Eaa}), onto the 2D plane using multi-dimensional scaling algorithm. Each of the four plot shows one of the ways to group amino acids based on various biological properties. The top left plot shows the hydrophobic (amino acids encircled by the grey area) versus hydrophilic (all the remaining amino acids). The top right plot shows the clustering based on size: the larger amino acids are highlighted by the grey area, while the amino acids outside the region have the smaller size. Bottom left shows the group of aromatic amino acids. The bottom right plot shows the cluster of negatively and positively charged amino acids. }
\label{fig:clust4}
\end{figure}

From \eqref{eq:emb}, the amino acid embeddings $E_{aa} \in \mathbb{R}^{|V_t|\times d}$ can be defined as 
\begin{align}
E_{aa} = \theta E.
\label{eq:Eaa}
\end{align}
We use multi-dimensional scaling (MDS) algorithm to project all the 20 amino acids $E_{aa}$ from $d$-dimensional space into 2D. The resulting scatter plot is shown in Fig. \ref{fig:clust4}. Each of the four plot shows one of the ways to group amino acids based on various biological properties. The top left plot shows the hydrophobic (amino acids encircled by the grey area) versus hydrophilic (all the remaining amino acids). The top right plot shows the clustering based on size: the larger amino acids are highlighted by the grey area, while the amino acids outside the region have the smaller size. Bottom left shows the group of aromatic amino acids. Finally, the bottom right plot shows the cluster of negatively and positively charged amino acids.  As can be seen, the learned embedding matrix for amino acids encodes meaningfull grouppings of the amino acids, with the amino acids from the same biological group collocated close to each other. This signifies that ReprogBert is able to properly project pretrained English word token embeddings and create new protein embeddings with meaningful biological properties. 

Since $x_s$ is usually a dense vector, its projection onto embedding vector has limited interpretability in that all rows of $E$ are mixed during the projection. To better understand and visualize the mapping between amino acids and English tokens, we propose to replace the above projection with cross-attention mechanism. 

\subsection{Cross-Attention}
\label{sec:apendix_crs_atn}

For this, we redefine $\theta$ to be the amino acids embedding matrix and keep the English tokens embedding matrix as is:
\begin{align}
\theta &\in \mathbb{R}^{|V_t|\times d}\\
E &\in \mathbb{R}^{|V_s|\times d},
\end{align}
and define the multi-head cross attention mechanism between $\theta$ and $E$. We split $d$ (in our case $d=768$) into $h$ parts of size $d_i=d/h$ each, resulting in
\begin{align}
\theta &= \left[\theta_1, \ldots, \theta_h\right], \quad \text{for}~\theta_i \in \mathbb{R}^{|V_t| \times d_i} \\
E &= \left[E_1, \ldots, E_h\right], \quad \text{for}~E_i \in \mathbb{R}^{|V_s| \times d_i},
\end{align}
and also define $d_i$ learnable projection matrices
\begin{align}
W_i^{Q} \in \mathbb{R}^{d_i\times d_i} \\
W_i^{K} \in \mathbb{R}^{d_i\times d_i} \\
W_i^{V} \in \mathbb{R}^{d_i\times d_i}.
\end{align}

The projection of $\theta$ and $E$ gives us the key, query and value matrices:
\begin{align}
Q_i &= \theta_i W_i^{Q}  \\
K_i &= E_i W_i^{K} \\
V_i &= E_i W_i^{V},
\end{align}
which are then projected to obtain the cross-attention
\begin{align}
\bar{E} &= \left[\bar{E}_1, \ldots, \bar{E}_h\right], \quad \text{where}~\bar{E} \in \mathbb{R}^{|V_t|\times d}\\
\bar{E}_i &= hardmax\left(\frac{Q_iK_i^T}{\sqrt{d_i}}\right)V_i, \quad \text{where}~\bar{E}_i \in \mathbb{R}^{|V_t|\times d_i},
\end{align}
where to compute the attention weights, we replaced the traditional softmax operation with hardmax, which converts the probabilities into one-hot prepresentation  
\begin{align}
A_i = hardmax\left(\frac{Q_iK_i^T}{\sqrt{d_i}}\right) \in \mathbb{R}^{|V_t| \times |V_s|},
\label{eq:mapA_i}
\end{align}
where each row is all zeros except a single one in one of the columns. In practice, to enable differentiation, we use straight-through Gumbel-softmax to implement the differentiable hardmax operation.

Note that by using one-hot attention enables us to see exactly which of the amino acids is associated with which English token, therefore improve mapping visualization. However, since we use multiple heads, the mapping is not one to one but can be one to many, where a single amino acid is mapped into multiple English tokens due to the concatenation of resulting sub-embedding matrices:
\begin{align}
\bar{E} &= \left[\bar{E}_1, \ldots, \bar{E}_h\right].
\end{align}

Now the the input amino acids sequence can be directly used to as input to the model:
\begin{align}
x_s^{\bar{E}} = x_s\bar{E},
\end{align}
and continue the usual processing through the transformer layers and blocks. Note that similar derivations can be constructed to derive the cross-attention for $\gamma$, mapping amino acid space into the English token space.

To summarize, the projection method in Section \ref{sec:apendix_proj} optimizes only two learnable matrices $\theta$ and $\gamma$, while keeping the rest of the ReprogBert parameters frozen. On the other hand, in the cross-attention approach, additionally to $\theta$ and $\gamma$, we also learn $h$ attention matrices $W_i^{Q}$, $W_i^{K}$, and $W_i^{V}$.

The comparison between the projection and cross-attention approaches is shown in Table \ref{tab:crsattn}.

\begin{table}[]
\centering
\resizebox{0.5\textwidth}{!}{%
\begin{tabular}{@{}lccc@{}}
\toprule
\multicolumn{1}{c}{}                  & \multicolumn{3}{c}{SabDab CDR-H3} \\ \midrule
\multicolumn{1}{c}{}                  & PPL-ProGen    & AAR     & DIV     \\ \midrule
ReprogBert (projection)               & 5.4           & 32.6    & 67.4    \\
ReprogBert (cross-attention, 1 head)  & 94.5          & 5.2     & 94.1    \\
ReprogBert (cross-attention, 2 heads) & 94.1          & 5.0     & 93.2    \\
ReprogBert (cross-attention, 4 heads) & 7.0           & 26.8    & 79.5    \\
ReprogBert (cross-attention, 6 heads) & 5.8           & 30.2    & 68.2    \\
ReprogBert (cross-attention, 8 heads) & 5.9           & 31.6    & 67.9    \\
ReprogBert (cross-attention, 12 heads) & 5.9          & 31.1    & 66.7 \\ \bottomrule
\end{tabular}%
}
\caption{Performance comparison between various mappings in ReprogBert as trained and tested on SabDab CDR-H3 dataset. Cross-attention with a single or 2 heads is not performing well, while larger number of heads improves the performance. 8 heads cross-attention matches the performance of linear projection (first row).}
\label{tab:crsattn}
\end{table}

\begin{figure}[!ht]
\centering
\includegraphics[width=0.5\linewidth]{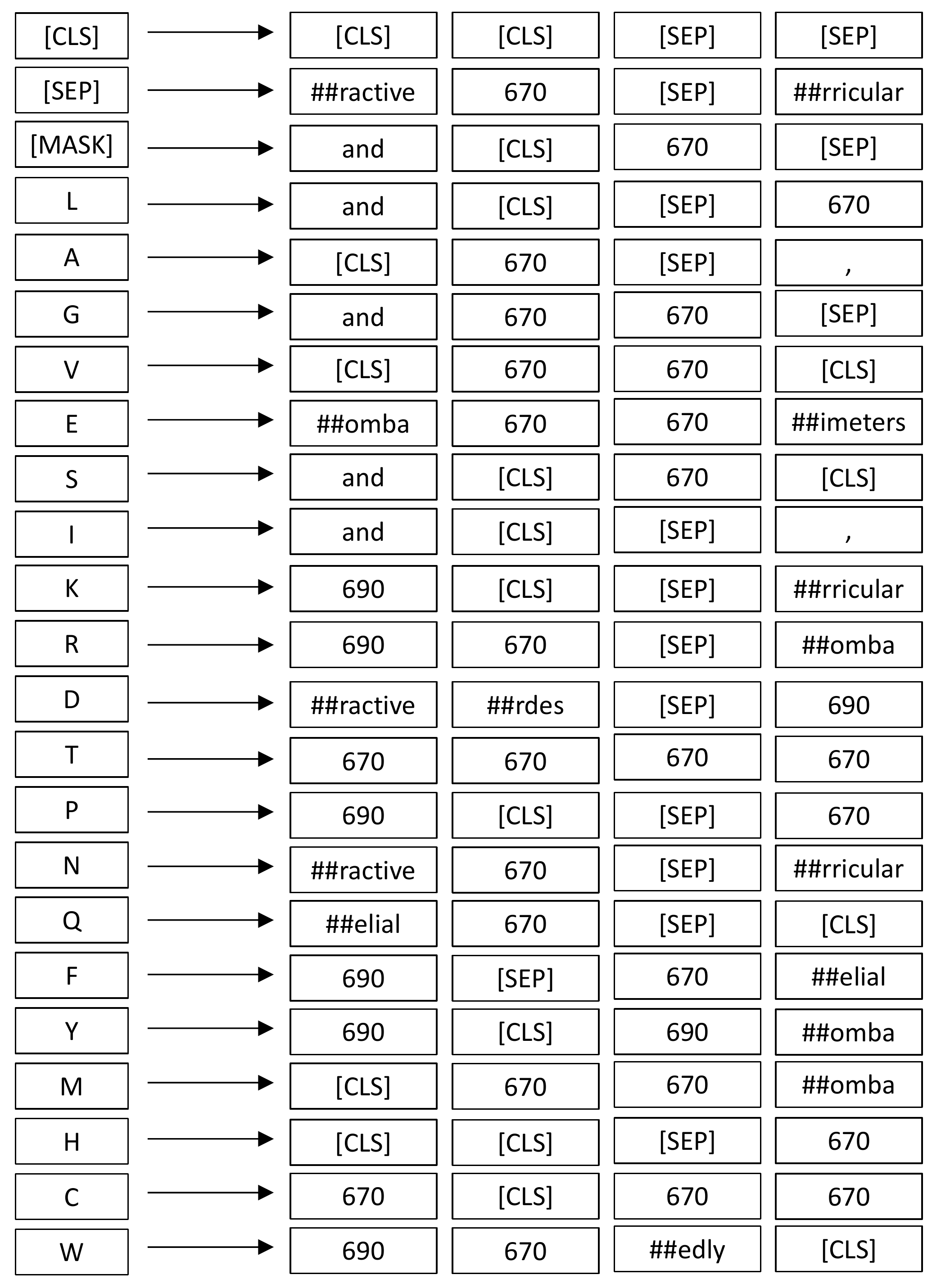}
\caption{The mapping ($A_i$ as defined in equation \eqref{eq:mapA_i}) between amino acid vocabulary and the English word tokens for ReprogBert model with cross-attention using 4 heads. As can be seen, each amino acid is mapped into some combination of 4 word tokens, where each of 4-token combindations is unique. It appears there is no particular semantic meaning in this map, rather it follows some statistical relationships the model optimized during the training. At the same time, it is interesting to observe that to get a good performance the model had to use at least 4 heads (corresponding to 4 word tokens per amino acid), while using fewer, such as a single or two words per token, did not produce satisfactory performance.}
\label{fig:map4}
\end{figure}

\begin{figure}[!ht]
\centering
\includegraphics[width=0.99\linewidth]{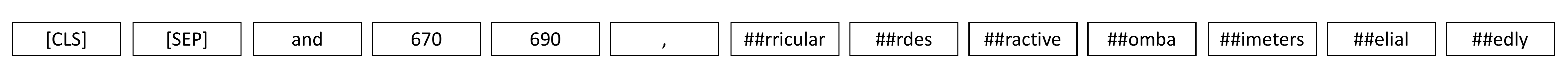}
\caption{All the word tokens used in the map of Figure \ref{fig:map4}. There are in total 13 word tokens.}
\label{fig:mapvocab}
\end{figure}

As, can be seen the cross-attention with 4 heads or more performs well, with 8 and 12 heads matching the projection method closely. In Figure \ref{fig:map4} we present the map for the case of cross-attention with 4 heads for the ease of visualization. Here, each amino acid is mapped into some combination of 4 word tokens, where each of 4-token combindations is unique. The set of all the word tokens (13 of them) which were used in the map are shown in Figure \ref{fig:mapvocab}. 

\FloatBarrier
\section{Recovery and Diversity Metrics}
\begin{figure}[!ht]
\centering
\includegraphics[width=0.99\linewidth]{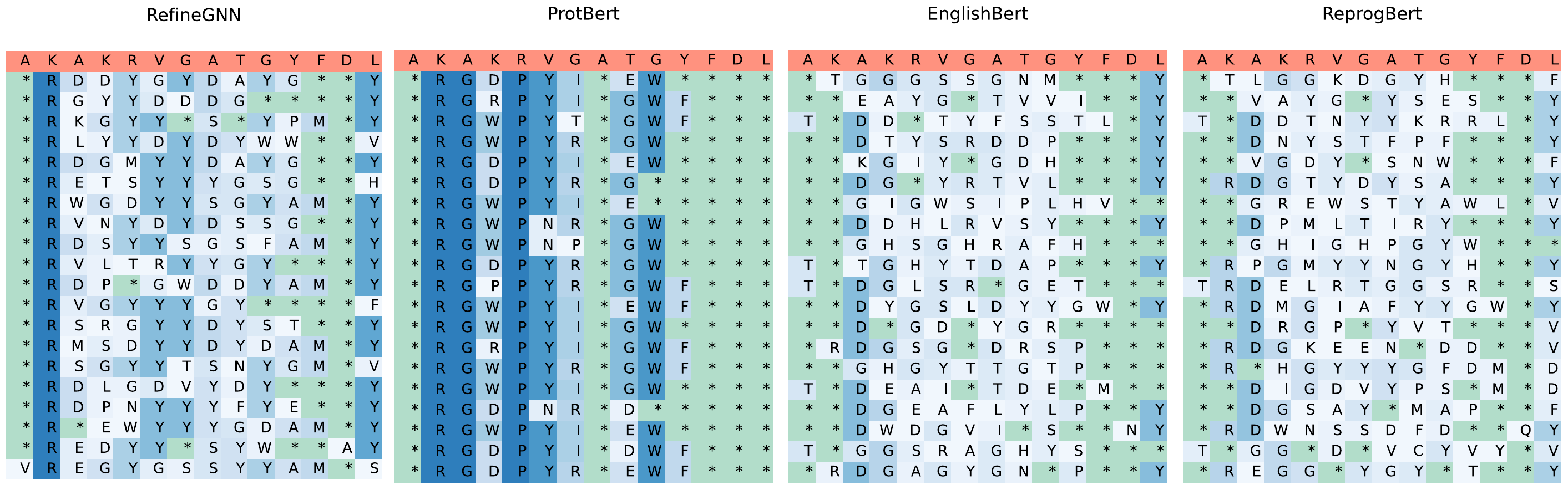}
\caption{Visualization of the recovery and diversity metrics for CDR-H3 (PDB ID 2r56) across different models. The top red line shows the ground truth CDR-H3 sequence, while the next lines show the generated CDR-H3 by each of the model. The green cell with the star symbol represents the same amino acid as in the ground truth, while the blue cell shows new and different generated residues. The shade of the blue cell represents the frequency of the amino acid in that column. We see that ReprogBert has highest diversity represented by the largest number of light blue cells, at the same time ProtBert has most green cells (highest AAR), but also many dark blue cells (low diversity). It can also be seen that RefineGNN has lower diversity and lower recovery, as compared to ReprogBert.}
\label{fig:2r56}
\end{figure}

\begin{figure}[!ht]
\centering
\includegraphics[width=0.99\linewidth]{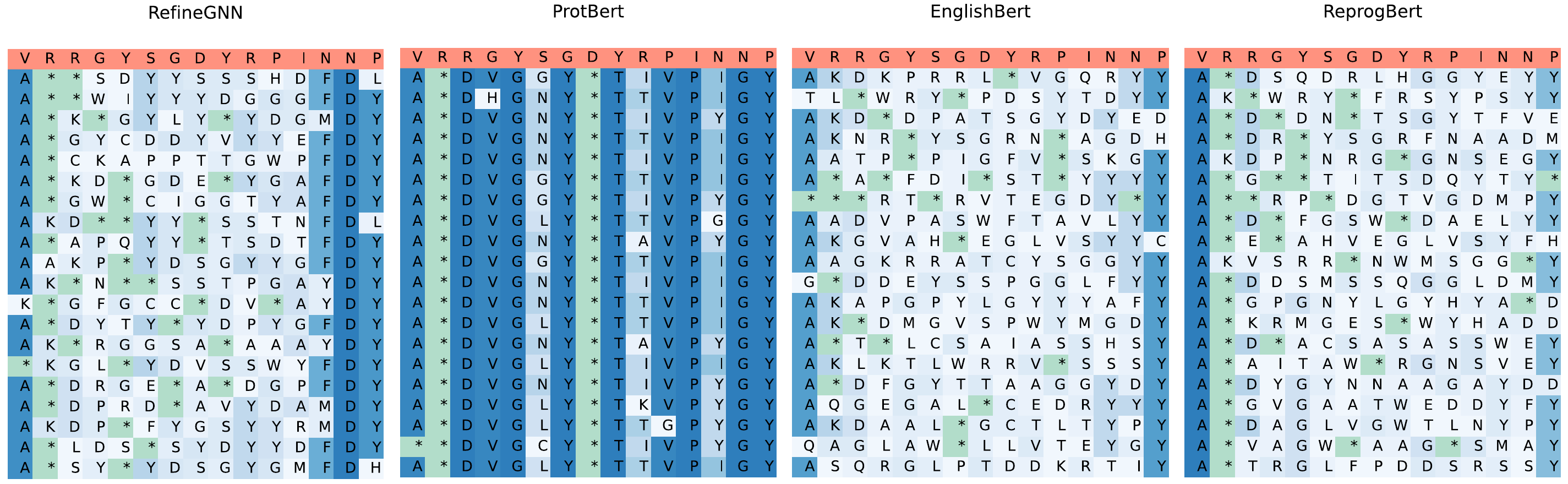}
\caption{Visualization of the recovery and diversity metrics for CDR-H3 (PDB ID 5y7z) across different models. The top red line shows the ground truth CDR-H3 sequence, while the next lines show the generated CDR-H3 by each of the model. The green cell with the star symbol represents the same amino acid as in the ground truth, while the blue cell shows new and different generated residues. The shade of the blue cell represents the frequency of the amino acid in that column. We see that ReprogBert has highest diversity represented by the largest number of light blue cells, at the same time ProtBert has most green cells (highest AAR), but also many dark blue cells (low diversity). It can also be seen that RefineGNN has lower diversity and lower recovery, as compared to ReprogBert.}
\label{fig:5y7z}
\end{figure}

Figures \ref{fig:2r56} and \ref{fig:5y7z} show additional visualizations of the recovery and diversity metrics for CDR-H3 across different methods.

\FloatBarrier
\section{Physicochemical Property Comparison}
\begin{figure}[!ht]
\centering
\includegraphics[width=0.99\linewidth]{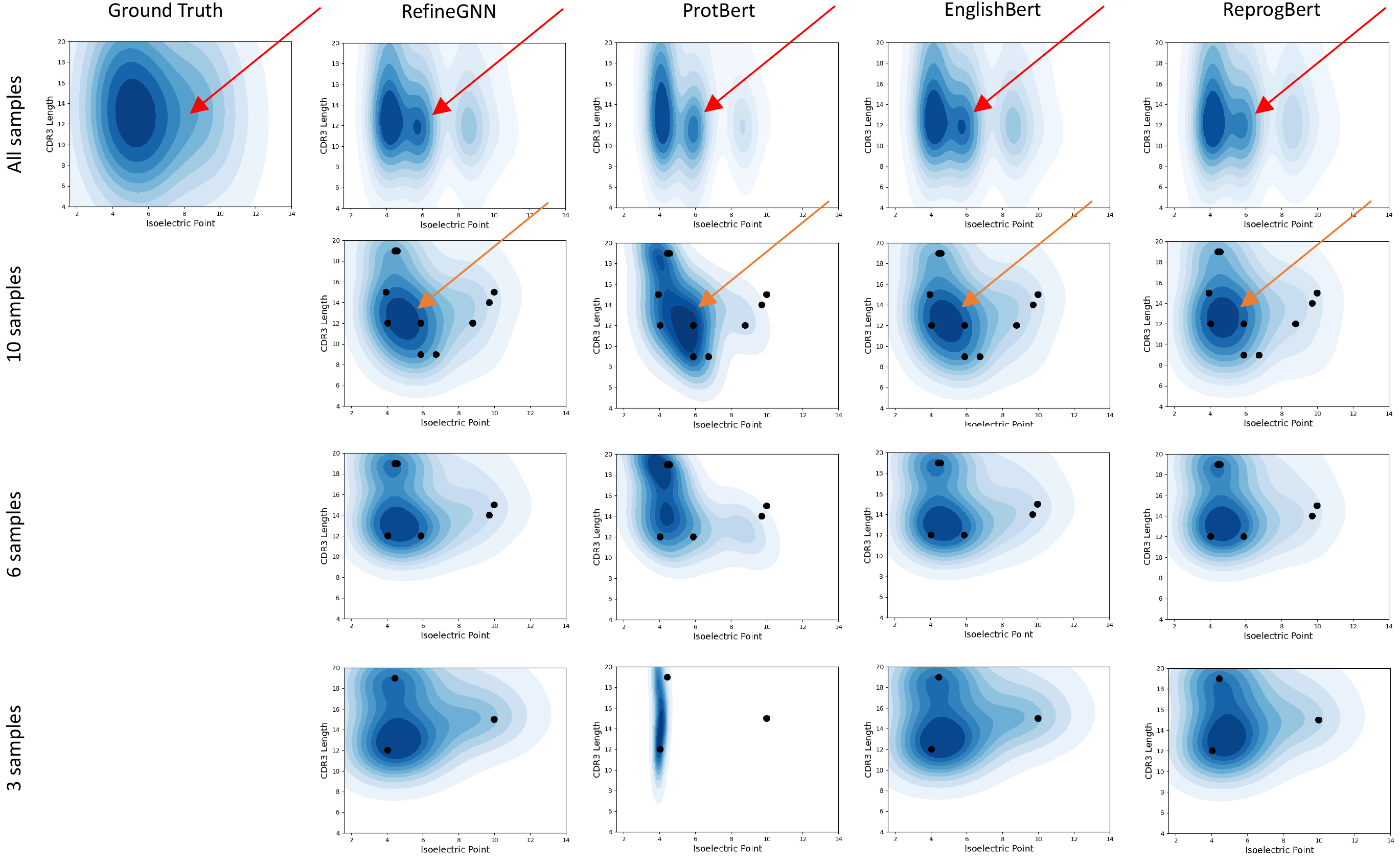}
\caption{2D kernel density plot as a function of isoelectric point, the pH at which a particular molecule carries no net electrical charge, and length of generated CDR-H3 sequences on the test set of SabDab dataset. Black dots indicate the ground truth CDR-H3. The top row shows the density of all the sequences, while the following rows show the density for 10, 6 and 3 samples. It can be seen that the ground truth density of all the sequnces (top left corner) has one pronounced peak (for the CDR3 length 13 and pH 5) and another smaller increase of density marked with red arrow. Comparing this region across other models, we see that ReprogBert has the closes resemblance to the ground truth, while others place too much weight there. The second row from the top shows the density for 10 protein sequences, where visual inspection of the region marked with orange arrow reveals that ReprogBert has closest similarity to the ground truth based on the distribution and orientation of the highly dense region, while for other methods theshape of this region is tilted and a second minimum appears. }
\label{fig:isoelectric}
\end{figure}

Finally, in Figure \ref{fig:isoelectric} we show 2D kernel density plot as a function of isoelectric point and length of generated CDR-H3 sequences on the test set of SabDab dataset.

\end{document}